\pdfoutput=1

\documentclass[11pt, reqno]{article}
\usepackage{jheppub}
\usepackage{epsfig}
\usepackage{amssymb}
\usepackage{amsmath}
\usepackage{mathrsfs}
\usepackage{hyperref}
\usepackage{multirow}

\def\MHV{\textrm{MHV}}

\def\OO{\mathcal{O}}
\def\be{\begin{equation}}
\def\ee{\end{equation}}
\def\ba{\begin{eqnarray}}
\def\ea{\end{eqnarray}}
\def\l{\langle}
\def\r{\rangle}

\def\nl{\nonumber\\}

\def\SS{\mathcal{S}}

\def\Eq#1{Eq.~(\ref{#1})}
\def\Li{\textrm{Li}}
\def\ibar{\bar{i}}
\def\jbar{\bar{j}}
\def\kbar{\bar{k}}

\title{Superconformal symmetry and two-loop amplitudes in planar N=4 super Yang-Mills}
\author[a,b]{S. Caron-Huot}
\affiliation[a]{School of Natural Sciences, Institute for Advanced Study, Princeton, NJ 08540, USA}
\affiliation[b]{Also at Kavli Institute for Theoretical Physics, Santa Barbara, CA 93106-4030, USA}
\emailAdd{schuot@ias.edu}
\abstract{Scattering amplitudes in superconformal field theories do not enjoy this symmetry,
because the definition of asymptotic states involve a notion of infinity.
Concentrating on planar $\mathcal{N}=4$ Yang-Mills, we consider a generalization of scattering amplitudes which depends on twice as many Grassmann variables.
We conjecture that it restores at least half of the superconformal symmetries, and all of the dual superconformal symmetries.
The object arises naturally as the dual of a null polygonal Wilson loop in an $(x,\theta,\bar\theta)$ superspace.
We support the conjecture by using it to obtain the total differential of all $n$-point two-loop MHV amplitudes,
and showing that the result passes consistency checks.  Potential all-loop constraints are also discussed.}
\arxivnumber{1105.5606 [hep-th]}

\begin{document}
\maketitle

\section{Introduction}
\label{sec:intro}

Planar $\mathcal{N}=4$ super Yang-Mills is integrable.  This means that this theory will be solved exactly.
Recent advances on two-loop scattering amplitudes \cite{volovichetal,DDS,khoze2d,duhr2d}
lend renowed credence to the long standing
hope, that compact analytic formulas describing nontrivial multiloop scattering processes can be found \cite{BDS}.

Compared to the now relatively mature problem of the spectrum of anomalous dimension of local operators, the problem of computing the S-matrix
is still rather in its infancy.  In the spectrum problem, integrability was the key to unlocking
the theory \cite{minahanzarembo,beisertstaudacher}.  Integrability made it possible to calculate,
to any desired accuracy, quantities with a nontrivial dependence on the t'Hooft coupling $\lambda=g^2_{\textrm{YM}}N_c$
for arbitrary finite value of the coupling \cite{BES,BDS0,Ysystem,intreview}.   On the other hand, the recent advances
in S-matrix computations did not really exploit integrability, beyond the duality with Wilson loops which was used to afford some simplifications.
The methods used, ranging from physical intuition regarding collinear limits and Regge limits, to an important innovation
in the handling of transcendental functions called the ``symbol'', are not specific to integrable theories.

The fact that integrability was not central should be viewed as a good thing, as this suggests
a potential for leaking these developments beyond the realm of $\mathcal{N}=4$.
On the other hand, it appears likely that a good grasp of the symmetries
will become increasingly necessary as we progress toward the depths of this theory.
Contact with integrability in scattering amplitudes was made from the string side at strong coupling \cite{smatrixYsystem}.
Arguably, it still remains elusive at weak coupling, despite at best some glimpses at it through the OPE approach to computing Wilson loops \cite{OPE1,OPE2,OPE3}.
There may also some constrains, most notably Regge limits \cite{Bartels1,Bartels2} or self-crossing relations \cite{Dorn},
which although not obviously related to the Yangian symmetry, maybe have not yet been fully exploited.

Yangian symmetry at loop level appears subtle.
Most generators of the Yangian are known to require modification at loop level, including the ordinary superconformal
and dual superconformal symmetry. They acquire some right-hand side when acting on amplitudes:
\be
 \sum_j \theta_j \frac{d}{dx_j} A_n(x_i,\theta_i) = \delta A_n  \label{RHS}
\ee
Here $x$ and $\theta$ are the so-called dual coordinates, explained below.
This equation is not an ``anomaly,'' in the sense that the right-hand side $\delta A_n$ can be absorbed into
an appropriate modification of the symmetry generators on left-hand side, without changing the algebra.  The deformation is well-understood at
1-loop \cite{anomaly1,anomaly2,anomaly3,anomaly4}, but in principle a new formula has to be re-derived at every new loop order.
We stress that the problem is deeper than infrared divergences.
The origin of the problem, as understood in these works, lies in the notion of an asymptotic particle state prepared at infinity not being conformally invariant.
A conformal transformation will cause the external state to radiate some collinear particles.
This nontrivial affects even canonical infrared finite quantities such as the ratio of NMHV to MHV amplitudes \cite{anomaly1}.
In fact, as stressed in \cite{abcct}, the whole fact that the MHV remainder function in $\mathcal{N}=4$ is not constant is such an effect.

In this paper we propose enlarging scattering amplitudes into a bigger object.  The idea is that within this bigger class of object,
no right-hand sides would appear and the naive form of the generators should be the correct one, at least for some part of the Yangian algebra.
Specifically, we propose that a generalization of scattering amplitudes exists, that depends on $4n$ new Grassmann variables $\bar\chi_i$,
which can be viewed as constrained $\bar\theta's$, such that the dual superconformal symmetries become
\be
 \left( \sum_j \theta_j \frac{d}{dx_j} + \sum_j \frac{\partial}{\partial \bar\theta_j}\right)A_n(x_i,\theta_i,\bar\theta_i) = 0
\ee
for any infrared finite observable, such as the MHV remainder function or the ratio function.

The motivation for this new object comes from the Wilson loop side of the scattering-amplitude/Wilson loop duality.
MHV scattering amplitudes are dual to bosonic null polygonal Wilson loops \cite{aldaymalda,WL1,WLDrummond2,WLDrummond},
and other helicity amplitudes are dual to supersymmetric null polygon Wilson loops \cite{masonskinner,wilsonloop1,EKS2,reg1,reg2}.
While the duality requires only a Wilson loop in a chiral half $(x_i,\theta_i)$ of the superspace,
the proposed generalization is a Wilson loop in the full $(x_i,\theta_i,\bar\theta_i)$ $\mathcal{N}=4$ superspace.
The previously missing half of the variables, restores the chiral half of the dual supersymmetries which was broken.\footnote{There seems to be
no obvious connection with the recent non-chiral formulation of scattering amplitudes \cite{yutin}.}.

As a main application, we will see that the proposal leads, when combined with explicit component expressions for the super-Wilson loop,
to a concrete and remarkably powerful method (by today's standards; hopefully, less remarkable by tomorrow's standards)
for computing any derivative of a multiloop scattering amplitude.
The main result of this paper is an integral representation for the total differential of the two-loop MHV $n$-point amplitude, and its symbol; for $n\geq 7$ this was not previously known.

This paper is organized as follows. In section \ref{sec:prelim} we review the basic notation and formalism we will be using.
This includes a six-dimensional notation recently developed by Weinberg, which can be useful in an arbitrary conformal field theory,
as well as a brief review of symbols.  In section \ref{sec:susy} we state a precise conjecture regarding the supersymmetric Wilson loops,
and work out its immediate implications for MHV amplitudes.
In section \ref{sec:main} we apply it to a general two-loop $n$-point MHV amplitudes and explain how it leads to a formula
for the derivative of this object. Section \ref{sec:checks} describes several consistency checks which we have applied to this result.
We finish with some conclusions.  The symbol of the $n$-point amplitude is reproduced in Appendix \ref{app:symbol}, and in two other appendices
we describe a general method for computing the symbol of a rational integral as well as the reduction to two-dimensional kinematics.

\section{Notations and preliminaries}
\label{sec:prelim}

\subsection{Six-dimensional notation}

In this paper we will be will be concerned with finite quantities in a conformal field theory, and expect conformal invariant answers.
In this section we would like to describe a formalism which allows to maintain conformal invariance manifest at every stage, in any conformal field theory,
following a recent suggestion by Weinberg \cite{weinberg}.  We found it quite useful in practice.

We will be working in coordinate space.  The usual spacetime coordinates $x^\mu$ ,where $\mu=1\ldots4$, single out  a preferred
Poincar\'e subgroup of the SO(4,2) conformal group.  To avoid singling out such a subgroup, it is advantageous to
package the coordinates into a null 6-vector
\be
 x^\mu \to X^i\equiv \left(\begin{array}{c} x^\mu \\ X^+ \\ X^-\end{array}\right) = \left(\begin{array}{c} x^\mu \\ 1 \\ x^2 \end{array}\right) \label{6dreps}
\ee
where $X$'s are identified modulo rescaling, $X\simeq \alpha X$.
The six-dimensional coordinates $X$ modulo rescaling can be interpreted, in a standard way, as parametrizing anti-de-Sitter space.
The null condition $X^2=0$ then gives the usual realization of (conformally compactified) four-dimensional Minkowski space with the boundary of AdS space
(see, for instance, \cite{wittenAdS} for a pedagogical presentation).  The conformal field theory and its fields are functions on this four-dimensional boundary.
Thus we are simply using convenient coordinates to describe Minkowski spacetime.

A procedure for describing general spin fields in this language was given in \cite{weinberg}.
We will be using a slight variation of this method.
The simplest example is a scalar field with mass dimension $1$, which becomes a function $\phi(X)$ with homogeneity degree $-1$ in $X$.
Its two-point function, up to normalization, is then the only invariant with the correct weight
\be
 \l \phi(X)\phi(X)\r= \frac{1}{\l XY\r},
\ee
where $\l XY\r$ is the six-dimensional dot product.
As in \cite{weinberg}, we will need to lift 2-component 4-dimensional Weyl spinors $\psi_\alpha$,
to 4-component 6-dimensional Weyl spinors $\psi_a$. We will do this in a slightly different way, though.
We choose to remove the extra two components through the equivalence relation
$\psi \simeq \psi + X\tilde\zeta$, where the second term involves gamma matrix multiplication.
Because $X^2=0$, the space of spinors of the form $X\tilde\zeta$ has dimensionality two as required.
In general a field with $s$ spinor indices and dimensionality $d$ will be mapped to a field with homogeneity degree $-(d{+}\frac12 s)$.
Thus, a canonical fermion field will have the propagator\footnote{Our fermions are related simply to those of Weinberg \cite{weinberg},
which satisfy the constraint $X\tilde\psi_\textrm{Weinberg}=0$: $\tilde\psi_\textrm{Weinberg} = X\psi_\textrm{us}$.}
\be
 \l \psi_a(X) \tilde\psi^b(Y)\r = \frac{\delta_a^b}{\l XY\r^2}.
\ee

Our notation is closely related to the ``momentum twistors'' of Hodges \cite{hodges}.  If $X$ is a null six-vector,
the antisymmetric matrix $X^{ab}$ has rank two and can be written
\be
 X^{ab}= X^{a\,(1)}\wedge X^{b\,(2)}
\ee
for some four-component ``momentum twistors'' $X^{a\,(1)},X^{a\,(2)}$.  These are the two solutions to the Weyl equation $X\psi^a=0$.

This is the usual twistor correspondence between a point in spacetime and the line $(X^{(1)},X^{(2)})$
in momentum twistor space. (This is the standard twistor correspondence.
We use the name ``momentum twistor'', as opposed to simply twistors, only as a reminder that the
spacetime we are working with, on which the Wilson loop is defined, corresponds physically
to momentum space in the scattering amplitudes side.)

The momentum twistor variables are the most efficient way to parametrize a null polygon contour, or, equivalently, the data of a massless planar scattering S-matrix element.
Let us quickly review the mapping, due to Hodges \cite{hodges}.  The external momenta $p_i$ are related to the dual coordinates $x_i$ through
\be
 p_i^\mu = x_i^\mu-x_{i{-}1}^\mu,
\ee
where $x_{n{+}1}$ is periodically identified with $x_1$.  Each point $x_i$ is \emph{a-priori} associated with two momentum twistors, but
the null condition $(x_i-x_{i{-}1})^2=0$ means that $x_i$ and $x_{i{-}1}$ share a momentum twistor;  call it $Z_i$.  Therefore
the null polygon is described by $n$ free momentum twistors $Z_i$, and its corners are located at $X_i=Z_i\wedge Z_{i{+}1}$.
The same can be done supersymmetrically.  Writing $p_i^{\alpha\dot\alpha}=\lambda_i^\alpha \tilde\lambda_i^{\dot\alpha}$ for spinors $i$,
and using Grassmann variables $\eta_i^A$ to describe the external polarizations, one introduces $\theta$'s through
\be
 \lambda^\alpha_i\eta_i^A = \theta_i^{\alpha A} - \theta_{i{-}1}^{\alpha A}.
\ee 
The momentum supertwistors are then, explicitly, $(Z_i,\chi_i) = (\lambda_i,(\lambda_i x_i),(\lambda_i\theta_i))$.

The dot product between two 6-vectors is precisely the fully antisymmetric contraction of the four momentum twistors
$ \l XY\r = \l X_a^{(1)}X_a^{(2)}Y_a^{(1)}Y_a^{(2)} \r$,
which explains the bracket notation.
We will repeatedly write strings of six-vectors acting on a momentum twistor, for instance objets like $\l\theta XYW\theta'\r$.
If $Y=Y^{(1)}\wedge Y^{(2)}$, this is to be interpeted as
\be
 \l\theta XYW\theta'\r \equiv \l \theta XY^{(1)}\r\l Y^{(2)}W\theta'\r -\l \theta XY^{(2)}\r\l Y^{(1)}W\theta'\r.
\ee
These products obey the Clifford identity $\l \{ X,Y\} \r =-\l XY \r$.
The remaining rules needed to convert a general space-time expression to six-dimensional notation are:
$(x-y)^2\to \l XY\r/\l XI\r\l YI\r$, and $\frac{1}{\pi^2} \int d^4x \l XI\r^4 \to \int_X$, where $I$ is the infinity point.
The spacetime integral $\int_X$ is normalized so that
\be
 \Gamma(4)\int_{U}\frac{1}{\l U X\r^4} = \frac{1}{(\frac12\l X X\r)^2} \label{fundamentalint}.
\ee
Most integrals ever needed can be derived from this basic one or its derivatives, thanks to Feynman parameterization ---
a concrete example will be given in subsection \ref{sec:c2}.  

Derivatives $d/dX$ are well-defined modulo amounts proportional to $X$.
For instance, the following is the standard fermion kinetic term in the action
\be
 \frac{1}{\pi^2}\int \tilde\psi^{\dot\alpha}D_{\dot\alpha\alpha}\psi^\alpha \to
 \int_X \tilde\psi X \frac{\partial}{\partial X} \psi + \int_X \tilde\psi XAX\psi.  \label{fermionaction}
\ee
It is verified that this is invariant under the substitution $\tilde\psi\to \zeta X$.
Integration by parts, using the general identity $\int_X (X\frac{\partial}{\partial X} -2) F=0$, can be verified to relate this expression to a similar one with $\tilde\psi$ and $\psi$ exchanged, showing
that the unphysical $\psi$ polarizations also decouple. One has to use
the homogeneity
condition $\l X \frac{\partial}{\partial X}\r \psi=-2\psi$.

\subsection{Chalmers-Siegel action and gauge-fixing}

The Chalmers-Siegel formulation of gauge theory reorganizes perturbation theory as an
expansion around self-dual Yang-Mills. This arises naturally in the context of supersymmetric Wilson loops,
where supersymmetric Wilson loops computed within the self-dual theory are found to be dual to tree-level N${}^k$MHV tree-level scattering amplitudes \cite{masonskinner,wilsonloop1}.
This will also prove quite useful for the two-loop computation we present below.
The reorganization proceeds by rewriting the action as
\ba
 S_0 = \frac{1}{4g^2_\textrm{YM}}\int d^4x F_{\mu\nu}F^{\mu\nu} &\to& \frac{1}{4g^2_\textrm{YM}} \int d^4x F_{\alpha\beta}F^{\alpha\beta}
\nl &\simeq& \frac{N_c}{4\pi^2} \int d^4x F_{\alpha\beta} G^{\alpha\beta} -\frac{g^2_\textrm{YM}N_c^2}{(2\pi)^4} \int d^4x G_{\alpha\beta} G^{\alpha\beta}.
\label{Gsquare} 
\ea
The first line corresponds to the addition of a (imaginary) $\theta$-term, which has no effect in perturbation theory, while the second
line introduces an auxiliary self-dual field $G$.  
It is natural to perform similar manipulations with the gauge-fixing term in Feynman gauge:
\be
 S_\textrm{g.f} = \frac{1}{2g^2_\textrm{YM}}\int d^4x (\partial_\mu A^\mu)^2
 \simeq \frac{N_c}{4\pi^2} \int d^4x (\partial_\mu A^\mu) \lambda -\frac{g^2_\textrm{YM}N_c^2}{(2\pi)^4} \int d^4x ~\frac12\lambda^2. \label{lambdasquare}
\ee
These two equations are unified if we upgrade the spin-1 fields to general two-by-two matrices: 
\be
 G_{\alpha\beta} \to G_{\alpha\beta}^{\textrm{(symmetrical)}} +\frac12 \epsilon_{\alpha\beta} \lambda, \qquad F_{\alpha\beta} \to
 \partial_{\alpha\dot\alpha}A_\beta{}^{\dot\alpha} + [A_{\alpha\dot\alpha},A_\beta{}^{\dot\alpha}].
\ee
The propagator between $A$ and the new $G$ is simply
\be
 \l A_{\dot\alpha\alpha}(x) G_{\beta\gamma}(y)\r = \frac{(x-y)_{\dot\alpha\beta} \epsilon_{\alpha\gamma}}{(x-y)^4}.
\ee
This propagator is not conformally invariant.
Following the rules of the six-dimensional notation, $G$ is mapped to a weight $-3$ bispinor and the propagator becomes
\be
 \l A_\mu dx^\mu(X) (\theta G\theta')(Y)\r= \frac{ \l \theta XdXI\theta'\r}{\l XY\r^2 \l YI\r}
\ee
where we have contracted $G$ with two fiducial momentum twistors $\theta$ and $\theta'$ lying on the line $Y$.
The action of conformal transformations on the gauge-fixing
generates a four-dimensional family of gauges, labeled by different points $I$'s. (A five-dimensional extension where $I^2\neq0$ might also be consistent,
although we will not consider it.)  This freedom to choose $I$ will be exploited below. 

\subsection{Review of loop integrands}

Let us conclude these considerations with a simple example illustrating the formalism at work: the 1-loop correction to a bosonic Wilson loop (which is dual to a MHV amplitude).
To get the order $g^2_\textrm{YM}$ correction to the Wilson loop,
we need to bring down one power of the interaction Lagrangian  and take its correlation function with the Wilson loop.
First we need the correlation function between the Wilson loop and a single field strength at the point $Z$.\footnote{In this paper we will sometimes use the letter $Z$ without a subscript denotes a spacetime point; we hope this will cause no confusion with the momentum twistors $Z_i$.}
This receives one term per edge.
Parametrizing edge $i$ using $X=X_{i{-}1}+\tau_X X_i$, this gives
\be
 \l (\theta G\theta')(Z) W\r = \sum_i \l \theta X_{i{-}1} X_i I\theta'\r \int_0^\infty  \frac{d\tau_X}{\l Z(X_{i{-}1}+\tau X_i)\r^2\l ZI\r}.
\ee
The integral gives simply
\be
\frac{1}{\l Z i{-}1i\r\l Zii{+}1\r},
\ee
while the prefactor is simply $\l i{-}1ii{+}1\theta\r\l iI\theta'\r$.  Therefore that correlation function is
\be
 \l W (\theta G\theta')(Z)\r = \sum_i \frac{\l i{-}1ii{+}1\theta\r \l Ii\theta'\r}{\l Z i{-}1i\r\l Zii{+}1\r\l ZI\r}.  \label{inducedG}
\ee
The sum can be verified to be independent of $I$ using the four-term identity
\be
  \frac{\l i{-}1ii{+}1\theta\r \l Ii\theta'\r}{\l Z i{-}1i\r\l Zii{+}1\r\l ZI\r}
- \frac{\l i{-}1ii{+}1\theta\r \l I'i\theta'\r}{\l Z i{-}1i\r\l Zii{+}1\r\l Z'I\r}
= \frac{\l \theta ii{+}1II'\theta'\r}{\l Zii{+}1\r \l ZI\r\l ZI'\r}-\frac{\l \theta i{-}1iII'\theta'\r}{\l Zi{-}1i\r \l ZI\r\l ZI'\r}
\ee
which transforms it to a telescopic sum.
To obtain the loop integrand we substitute this into the interaction Lagrangian (the order $g^2_\textrm{YM}$ part of the Chalmers-Siegel Lagrangian),
which reads
\be
 S_\textrm{int}= -g^2 \frac12\int_Z G_{ab} Z^{ac}Z^{bd} G_{cd} + \ldots. \label{Gsquare2}
\ee
where $g^2=\frac{g^2_\textrm{YM}N_c}{16\pi^2}$ and the dots denote Yukawa and scalar self-interaction terms.
After using the simple identity $\l Zij\r \l Z (Ii)\cap (Ij)\r = \l Z(\ibar)\cap(\jbar)\r \l I ij\r \l ZI\r$
one obtains, directly\footnote{The intersection symbol is defined as $\l ij (abc)\cap(def)\r = \l iabc\r\l jdef\r - \l jabd\r\l idef\r$}
\be
 A_n^{\textrm{1-loop}} = -\sum_{1\leq i<j\leq n}  
 \int_Z \frac{ \l Z (i{-}1ii{+}1)\cap(j{-}1jj{+}1)\r\l I ij\r}{\l Zi{-}1i\r\l Zii{+}1\r\l Zj{-}1j\r\l Zjj{+}1\r\l ZI\r}.
\ee
This is exactly the 1-loop MHV integrand in the form found in \cite{abcct}.  The integrals will be discussed below.
The parity even part of this expression reproduces the more familiar sum over ``two-mass-easy'' box integrals \cite{bddk1} (this is verified in detail in \cite{EKS2}).

\subsection{Symbols}

Transcendental functions can typically be written in many equivalent ways, which usually makes it difficult to decide whether two expressions are equal or not.
A spectacular recent example is the equality between the compact formula given in \cite{volovichetal} and the original 17-page analytic formula obtained in \cite{DDS},
for the hexagon 2-loop remainder function.

The ``symbol'' of a transcendental function is a piece of invariant data which trivializes all identities between transcendental functions.
The symbol is uniquely defined and if two expressions have different symbols, one can immediately conclude that they differ.
The symbol is lossy, however, and from the equality of two symbols, one can generally only infer that they agree up to lower functional transcendentality, and choice
of Riemann sheet.

This subsection is meant as a lightning introduction to symbols and as a refreshment for the reader's memory.
Nice presentations can be found in \cite{volovichetal,OPE3}.
The transcendental functions of interest to us are (generalized) polylogarithms and are iterated integrals of the form
\be
 F(x) = \int_{x_0< x_1 < x_2 < x_3 < x} d\log X(x_1) d\log Y(x_2) d\log Z( x_3).  \label{iteratedintegral}
\ee
This particular integral would correspond to a transcendentality-3 function of $x$, e.g. some combination of $Li_3$, $\Li_2\log$ or $\log^3$ with various $x$-dependent
arguments.  The variable $x$ and the integration path
in general live in some multidimensional manifold; for us it is useful to visualize a point $x$ as a configuration of momentum twistors, or, equivalently, a collection of cross-ratios.
The symbol is simply the integrand of such an iterated integral, written as $ \log X \otimes \log  Y \otimes \log Z$, and abbreviated to
\be
 \SS[F(x)] = X\otimes Y \otimes Z.
\ee
The integrals of interest are such that they are contour-independent, expect for monodromy when the contour crosses a pole.
The upshot is that all identities between transcendental function descend to trivial identities at the level of the integrand,
and are mapped to simple algebraic identities on the symbol.  The only manipulations allowed are multilinearity, e.g.
\be
  X \otimes YW\otimes Z = X \otimes Y\otimes Z +  X \otimes W\otimes Z,
\ee
together with $\SS[\textrm{Constant}]=0$.  Frequent examples include: $\SS[\log x\log y] = x\otimes y + y\otimes x$ and $\SS[\Li_2(1-x)] = -x\otimes (1-x)$.

\section{Supersymmetric Wilson loops}
\label{sec:susy}

\subsection{Chiral Wilson loops revisited}

It is well-known that supersymmetric gauge theories can be interpreted as gauge theories in superspace.
Unbeknownst to the author of \cite{wilsonloop1} at the time of that work, is the equally well-known fact that 
$\mathcal{N}=4$ Yang-Mills can be interpreted as a gauge theory in a superspace $(x,\theta_\alpha^A,\bar\theta^{\dot\alpha}_A)$,
where $A=1\ldots 4$ are SU(4)${}_R$ indices \cite{witten86,harnadshnider}.  In that case, the superconnection is defined only on-shell.
In other words, given any on-shell solutions to the field equation, one can construct a superconnection which is supersymmetry covariant.
Actually, it descends from a superconnection in ten-dimensional $\mathcal{N}=1$ super Yang-Mills.

A chiral supersymmetric Wilson loop, for null polygon contours, was constructed in \cite{wilsonloop1} as follows.
The bosonic Wilson line along edge $i$ is translated into superspace:
\be
 \mathcal{P}e^{-\int_{i{-}1}^i A_\mu dx^\mu} \to e^{ c\chi_i^A \frac{\l *q_A\r}{\l * i\r}} \mathcal{P}e^{-\int_{i{-}1}^i A_\mu dx^\mu} e^{ -c\chi_i^A \frac{\l *q_A\r}{\l * i\r}},
\ee
where $c=(\frac{4\pi^2}{g^2_\textrm{YM}N_c})^{1/4}$.
This result is not quite yet supersymmetry covariant, e.g. not invariant under
$ c q^\alpha _A - \sum_i \lambda_{i\alpha} \frac{\partial}{\partial \chi_i^A}$.
The reason is that in attempting to verify invariance one needs to commute $q$ across the exponential,
and one picks up an anticommutator
\be
 \{ q_\alpha^A, q_\beta^B\} A_\mu = \epsilon_{\alpha\beta} D_\mu \phi^{AB}.
\ee
In \cite{wilsonloop1} this non-invariance by a total derivative
was canceled by inserting operators at the cusps, and the choice $\lambda_{*}=\lambda_{i{-}1}$ was made.\footnote{From this viewpoint we can
propose a better starting point for a nonperturbative definition: conjugate with
\be
 e^{ c \frac{\chi_i \l * i{+}1i{-}1 Q\r +\chi_{i{-}1} \l * ii{+}1 Q\r +\chi_{i{+}1} \l * i{-}1i Q\r}{\l *i{-}1ii{+}1\r}}
\ee
where $Q$ is the 4-component superconformal generator.  $Q$ will be slightly broken at the cusps by infrared regularization, but this can usually be controlled.
At least naively this definition removes cusp terms. The author thanks J.~Maldacena for related discussions.
}

Since the defining property of the object we are interested in is its supersymmetry covariance, and since an object with this property is obviously unique,
the result is guaranteed to be equivalent to the standard $\mathcal{N}=4$ superconnection with $\bar\theta$ set to zero in it,
up to equations of motion and gauge transformations.

We can even construct explicitly the super-gauge transformation which relates the two (a similar computation was carried out in \cite{BKS}).
From the integrability conditions obeyed by the standard superconnection,
$F_{(\alpha\beta)}^{(AB)}=0$ and $F_{(\alpha\beta)}^{A\dot\beta}=0$, where $(\alpha\beta)$ means symmetrical part, one concludes \cite{masonskinner} that the $\theta$-path
on top of edge $i$ can be deformed by any amount proportional to $\lambda_i$.   If we write the $\theta$-path as
\ba
 \theta(t) &=& (1-t) \theta_{i{-}1} + t \theta_i \nl
 &=& \chi_i \frac{-t \lambda_{i{-}1} \l ii{+}1\r + (1-t) \lambda_{i{+}1}\l i{-}1i\r}{\l i{-}1i\r\l ii{+}1\r}
   + f(t) \frac{\chi_{i{-}1} \lambda_i}{\l i{-}1i\r}
   - g(t) \frac{\chi_{i{+}1} \lambda_i}{\l ii{+}1\r}
  + h(t) \frac{\chi_i \lambda_i \l i{-}1\, i{+}1\r}{\l i{-}1i\r\l ii{+}1\r}  \nonumber
\ea
where $f(t)=(1-t)$, $g(t)=t$ and $h(t)=0$ are the canonical choices, what we are saying is that different functions $f(t)$, $g(t)$ and $h(t)$ are simply different gauge choices.
The gauge taken in \cite{wilsonloop1} corresponds to step-functions
$f(t)=\theta(\epsilon-t)$ and $g(t)=\theta(t-1+\epsilon)$ with $\epsilon$ very small.  This made manifest the fact
that $\chi_{i{-}1}$ and $\chi_{i{+}1}$ have no coupling inside to edge $i$. (In addition, the choice $h(t)=(1-t)\theta(t-\epsilon)$ was made in \cite{wilsonloop1}, which removed
the $\lambda_{i{+}1}$ component.  In retrospect, the choice $h(t)=0$ would have been much better, as it is conformally invariant.)

We conclude this discussion by reporting the beautifully simple expression we obtain for the abelian superconnection in 6-dimensional notation,
\be
 \mathcal{A} = \l AdZ\r + \l \tilde\psi \theta dZ \r + \l \theta d\theta \frac{\partial}{\partial Z}\r (\phi + \theta\psi + \theta G \theta).
\ee
up to signs and numerical factors ($\theta^a$ is a 4-spinor with $Z\theta=0$).

\subsection{A conjecture}

It is now easy to state our proposal, or conjecture:

\vspace{0.5cm}
{\bf Conjecture} There exists a generalization of planar scattering amplitudes, $A_n(x_i,\theta_i,\bar\theta_i)$,
such that
\begin{subequations}
\ba
          Q A_n &\equiv& \sum_i \left(\frac{\partial}{\partial \theta_i} + \bar\theta_i \frac{\partial}{\partial x_i}\right)A_n =0, \\
 \tilde Q A_n &\equiv& \sum_i \left(\theta_i\frac{\partial}{\partial x_i} + \frac{\partial}{\partial \bar\theta_i}\right)A_n =0,
\ea
\end{subequations}
plus similar equations for the dual special superconformal generators,
for any infrared-safe quantity, such as the remainder function or the ratio of two $A$'s with different MHV degree but at the same kinematic point.
Furthermore, this object should be equal to the expectation value of the supersymmetric Wilson loop of \cite{harnadshnider} on a null-polygonal Wilson loop contour.
\vspace{0.5cm}

To be fully precise, we should state what the ``null'' condition is, because it enters the data in $A_n$.
The condition should leave $3$ bosonic degrees of freedom per edge as usual,
plus four chiral fermions and four antichiral fermions, and there can be only one such superconformally invariant condition.
Tentatively, we would write it as  $p_i^2=0$, $p_{i\,\alpha\dot\alpha}(\theta_i-\theta_{i{-}1})^\alpha=0$,
$p_{i\,\alpha\dot\alpha}(\bar\theta_i-\bar\theta_{i{-}1})^{\dot\alpha}=0$ where $p_i=(x_i-x_{i{-}1}+\theta_i\bar\theta_{i{-}1}-\theta_{i{-}1}\bar\theta_i)$.
However we do not want to embark on checking these conditions here, because they will not be needed in this work, and present them only as a tentative guess.

A more important question, in our view, is to try to write these Ward identities in terms of unconstrained variables such as momentum twistors.
To first order in the new variables $\bar\chi$, the following algebra is easily deduced:
\begin{subequations}
\ba
 Q^a_A &\equiv& \sum_i Z_i^a \frac{\partial}{\partial \chi_i^A}, \\
 \tilde Q_a^A &\equiv& \sum_i \chi_i^A \frac{\partial}{\partial Z_i^a} + \sum_i (i{-}1ii{+}1)_a  \frac{\partial}{\partial \bar\chi_{iA}} + \ldots \label{generators}
\ea
\end{subequations}
where $a=1\ldots 4$.  This is the algebra we will be working with in this paper.
This does not look very parity symmetric, but is simply an artifact of working with chiral variables.
Unfortunately, we do not know yet a nice extension of this equation to higher orders in $\bar\chi$.\footnote{One promising possibility, suggested to the author by Beisert, Huang, Vergu and Skinner,
is to introduce two dual supertwistors $\mathcal{Z}$ and $\mathcal{W}$ subject to the superconformal constraints $\mathcal{Z}_i\cdot\mathcal{W}_i=0$,
$\mathcal{Z}_i\cdot\mathcal{W}_{i{-}1}=0$ and $\mathcal{Z}_i\cdot\mathcal{W}_{i{+}1}=0$.  The constraints can be used to solve for the bosonic $W$'s.  The symmetry generators are then
simply $\mathcal{Z}\partial/\partial \mathcal{Z} + \mathcal{W}\partial/\partial \mathcal{W}$.}

In terms of these variables, the first few terms of the super-Wilson loop are, in the normalization of \cite{wilsonloop1},
\be
 \mathcal{A} = dx_{\alpha\dot\alpha} (A^{\alpha\dot\alpha} + \tilde\psi^{\dot\alpha} \theta^\alpha + g^2\psi^{\alpha} \tilde\theta^{\dot\alpha} + \ldots),
\ee
where $\bar\theta_i = (\bar\chi_i \tilde\lambda_{i{+}1}-\bar\chi_{i{+}1} \tilde\lambda_{i})/[ii{+}1]$, or, in 6d notation
with $X=X_{i{-}1}+\tau_X X_i$,
\be
 \mathcal{A} = \l dX A\r + \chi_i \l i{-}1ii{+}1 \tilde \psi \r d\tau_X + \bar\chi_i \l i \psi\r d\tau_X + \ldots
\ee
The dots also contain a term $F\theta\bar\theta$ which would contribute to the same order in $\chi\bar\chi$.
Fortunately, we will manage to avoid needing this term.

The reader will notice that we have also not specified what the quantum definition of the Wilson loop should be
(e.g., including its proper regularization.\footnote{
 This does not mean an off-shell definition, which obviously would be too much to ask for in $\mathcal{N}=4$,
 but which is not required either.
 It should be perfectly legal to use equations of motion inside a correlation function
 provided $\delta^D(x)$ terms can be neglected, as is generally the case for Wilson loops.})  Only its bare expression is known.
Obviously, the defining property of the object we are interested is the Ward identities it should obey, and part of the conjecture
is that a regularization preserving this property exists.

We mention at least one, in all appearances natural, quantum interpretation of the classical Wilson loop which leads
to $\epsilon/\epsilon$-type violations of supersymmetry \cite{BKS}.  In other words,
regularization appears to be not entirely straightforward.
A regularization of the (square of) chiral Wilson loops based on correlation functions of BPS operators which corrects for these problems
was proposed thereafter in \cite{reg1,reg2}.  We will sidestep these questions in this paper by asking simply whether a consistent
answer obeying the Ward identities exists.
Normally, this is the right question to ask whenever one is worried about anomalies ---
we know of no example in quantum field theory where a consistent answer exists and yet there is an anomaly.

\subsection{MHV case}

We now concentrate on the MHV remainder function.  With no loss of generality we can expand
\be
 R_n^\MHV = \left.R_n^\MHV\right|_{\bar\chi,\chi=0} + \sum_{\substack{i,j \\ \textrm{non-adjacent}}} \frac{\bar\chi_i\chi_j}{\l i{-}1ii{+}1 j\r} C_{i,j}  + \textrm{(adjacent terms)} + \OO((\bar\chi\chi)^2). \label{expansion}
\ee
The ``non-adjacent'' terms cover all $i\neq j,j{-}1,j{+}1$.  For adjacent $i$ and $j$, we define $C_{i,j}=0$.

Let us work out the consequences of the Ward identities.
From $\frac{\partial}{\partial \bar\chi_i} \l i{-}1ii{+}1 Q\r R_n=0$
(no summation over $i$), we deduce the sum rule
\be
 \sum_j C_{i,j} =0. \label{sumrule}
\ee
Similarly, there is the parity conjugate sum rule
\be
 \sum_i C_{i,j} =0. \label{sumrule1}
\ee
From $\frac{\partial}{\partial \bar\chi_i} \l i{-}1i* Q\r R_n =0$, we deduce that
\be
 \frac{\partial^2}{\partial \bar\chi_i\partial\chi_{i{+}1}}R_n|_{\bar\chi,\chi=0} = \sum_j C_{i,j} \frac{\l i{-}1i* j\r}{\l i{-}1ii{+}1 j\r\l i{-}1ii{+}1*\r}
 \label{allsols}
\ee
which shows that all terms linear in $\chi$ and $\bar \chi$ are determined by the $C_{i,j}$'s.  For instance,
a similar equation using $\l i{-}1i{+}1*Q\r$ would determine the coefficient of $\bar\chi_i \chi_i$.
The right-hand side is independent of $*$, thanks to the sum rule (\ref{sumrule}).

From $\frac{\partial}{\partial \chi_j}\tilde QR_n=0$, we obtain that
\be
 \frac{\partial}{\partial Z_j^a} R_n|_{\bar\chi,\chi=0} = \sum_i (i{-}1ii{+}1)_a \frac{\partial^2}{\partial\bar\chi_i \partial\chi_j}R_n|_{\bar\chi,\chi=0}.
\ee
This shows that all bosonic first derivatives of the remainder function are determined by the order $\bar\chi\chi$ terms.  \Eq{allsols}
allows us to express the right-hand side in terms of the $C_{i,j}$'s solely.  Let us contract both sides of the previous equation with $dZ^a_j$ and sum over $j$.
Collecting the terms involving $C_{i,j}$ gives
\begin{align}
 \frac{C_{i,j}(i{-}1ii{+}1)_a}{\l i{-}1ii{+}1j\r\l i{-}1ii{+}1*\r} &\left( 
  \l i{-}1ii{+}1*\r dZ_j^a
 +\l i{-}1i*j\r dZ_{i{+}1}^a
 +\l i{-}1*i{+}1j\r dZ_{i}^a
 +\l *ii{+}1j\r dZ_{i{-}1}^a\right) \nl &= C_{i,j} \,d\log \l i{-}1ii{+}1j\r.
\end{align}
We have thus obtained a simple result for the total differential:
\be
  d  R_n|_{\bar\chi,\chi=0} = \sum_{i,j} C_{i,j} d\log \l i{-}1ii{+}1j\r.  \label{maindiff}
\ee
This is the main result of the present subsection.  The left-hand side is the total differential of the bosonic remainder function we are interested in.
The coefficients $C_{i,j}$ are defined by the expansion (\ref{expansion}). As explained in the previous subsection,
they may be computed by turning on fermions on specific edges of the Wilson loop.

\begin{figure}
\begin{center}
\includegraphics[width=4cm]{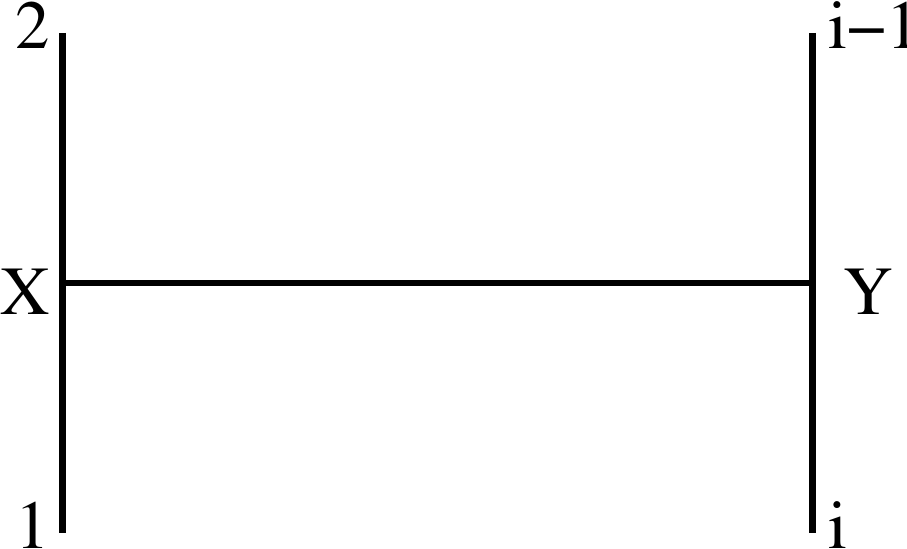}
\end{center}
\caption{The diagram giving the coefficient function $C_{2,i}$ at 1-loop.}
\label{fig:stretch1}
\end{figure}

Let us try to compute the $C_{i,j}$'s, say $C_{2,i}$, at one-loop.  We have a $\psi$ propagating along edge 2 and a $\tilde\psi$ propagating along edge $i$ (see Figure~\ref{fig:stretch1}),
joined by a propagator:
\be
 C_{2,i} = \int_0^\infty d\tau_X d\tau_Y \frac{\l \bar2i\r \l \ibar2\r}{\l XY\r^2} = \log u_{2,i{-}1,i,1} \label{1loopMHV}
\ee
where $X=(Z_1-\tau_X Z_3)\wedge Z_2$ and $Y=(Z_{i{-}1}-\tau_YZ_{i{+}1})\wedge Z_i$; we define cross-ratios as
\be
 u_{i,j,k,l} \equiv \frac{\l X_iX_j\r \l X_k X_l\r}{\l X_iX_k\r \l X_j X_l\r}.
\ee
We note three facts.  First, the sum rule \Eq{sumrule} is almost obeyed, the sum over $i$ being telescopic. However there are some uncanceled boundary terms.
Second, $C_{2,4}$ and $C_{2,n}$ are ill-defined, plagued with infrared divergences.
Third, leaving aside these boundary issues, the general term in \Eq{maindiff} is easily verified to perfectly match with the derivative of the 1-loop MHV amplitude \cite{bddk1}.

The slight mismatch here with boundary terms was to be fully expected: the naive Ward identities should be modified to account for the infrared divergences,
probably in a very similar way to the bosonic case \cite{conformalward}.  We note that there exists a canonical infrared regulator of the theory, the Coulomb branch,
in which these anomalies are removed by an appropriate action on the moduli space parameters \cite{fifthdimension}.  It would be tantalizing to interpret the $\bar\chi$ variables
as some sort of supersymmetrization of that regulator.
For the moment, we concentrate on the remainder function, which is infrared finite and for which the naive Ward identities are fully justified.

\subsection{Leading singularities}

The fact that MHV amplitudes turn out to be sums of pure transcendental functions (meaning, transcendental functions with rational prefactor equal to 1),
is generally understood from the fact that the only Yangian invariant with no fermion is a constant.
The reasoning takes its roots in generalized unitarity, and is based on the idea (highly plausible, but for which there exists no general proof at the moment) that
the coefficients of the transcendental functions should correspond to ``leading singularities'' of loop integrands, that is, by the loop integrals
performed on closed compact cycles (see, for instance, \cite{cachazoLS}).
These leading singularities have been proved to be Yangian invariant \cite{heslopLS,abcct}.

Therefore, a key question to ask about the $C_{i,j}$'s is, what are their leading singularities?  That is, what can arise if one performs
the integrals contributing to a $C_{i,j}$ on a closed compact cycle?

In the case of $\bar\chi=0$, all invariants are known to be closed contour integrals in the Grassmanian \cite{Grassmannian,Grassmannian2,Grassmannian3}.
In the MHV case, this meant the constant '1'.  The computation in the next section strongly supports the idea that the $C_{i,j}$'s enjoy similar properties --
their leading singularities at two-loops are the same as at one-loop, and are all '1'!
(The fact that one finds c-numbers, as opposed to rational functions, is the main statement.  However, we do not even find any $\frac12$.)
This immediately implies that the last entry of the symbol, at both one- and two- loops, is of the form $\l i{-}1ii{+}1j\r$.

It would be very interesting to see if this property is general -- whether leading singularities in the presence of $\bar\chi$'s at N${}^k$MHV for a given number of external legs
form a finite set to all loop orders --, 
and whether in the MHV cases these leading singularities are indeed saturated at 1-loop.\footnote{In the first version of this contribution, a symmetry argument was suggested based on an (incorrect) classification of ``invariants of homogeneous degree in $\bar\chi$ and $\chi$''.  Invariants with this property do not actually not exist when the full form of the superconformal algebra is used.  For this reason this discussion has been removed in second version. }.

\section{Two-loop MHV remainder function}
\label{sec:main}

We wish to use \Eq{maindiff} to determine
the derivative of the two-loop $n$-point remainder function.

\subsection{Outline of method}

There exists of course many different and equally valid ways to calculate the Feynman diagrams which arise.
We have employed a certain systematic method, which we now describe.  The advantage of the method
is that it automatically leads to relatively nice one-dimensional integrals, with manifest transcendentality degree 3.
In our view, this by far compensates for the somewhat more lengthy algebraic manipulations the method entails.
We also believe it likely that the method has a generalization to higher loop orders.

The steps of the method are the following:
\begin{itemize}
\item We fix the parameters $\tau_X$ and $\tau_Y$ that describe the insertion points of the fermions, and compute the integrand associated with a loop insertion point $Z$
using the method of Lagrangian insertion.
  This gives $C_{2,i}$ as a rational integral over 6 variables: $\tau_X$, $\tau_Y$ and the space-time point $Z$.
\item We perform the integration over $\tau_Y$.  The motivation for doing $\tau_Y$ first is the following.
The fact that the Lagrangian is chiral means that it has different types of singularities near $X$ and $Y$.
For instance, the OPE of the chiral Lagrangian with $\bar\psi$ and $\psi$ dictates double poles
in $\l ZY\r^2$ but only single poles in $\l ZX\r$.  Pleasingly, the integration
over $\tau_Y$ will simply convert the double poles to single poles, leaving a rational expression.  At this stage we thus have a five-dimensional rational integral with single poles.
\item We perform the four integrations over $Z$.  This has the structure of a standard loop integral and produces dilogarithms.
In fact, all the integrals will be known pentagons and scalar boxes.
\end{itemize}

We see that the problem is reduced to a one-dimensional integral over dilogarithms, without the need for further ``art'' or trickery.

We found the following technical tricks to be very useful:
\begin{itemize}
\item  We split the computation into four individually gauge-invariant and conformally invariant subsets.
This involves adding and subtracting certain ``geodesics in AdS'' edges to the polygon. While this may sound a bit unfamiliar, the idea in picture is very simple:
\be
\begin{array}{ccccccccc}
\raisebox{-0.7cm}{\includegraphics[scale=0.5]{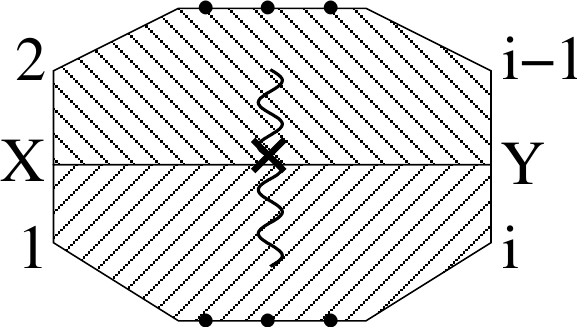} }&=&
\raisebox{-0.7cm}{\includegraphics[scale=0.5]{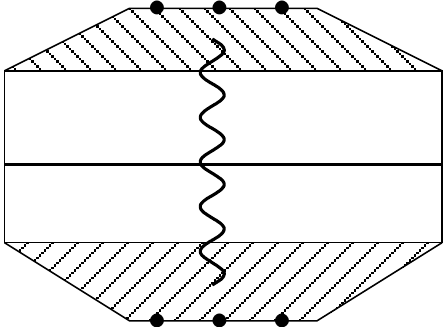}} &+&
\raisebox{-0.7cm}{\includegraphics[scale=0.5]{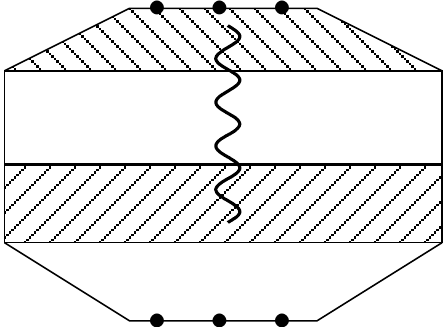}} &+&
\raisebox{-0.7cm}{\includegraphics[scale=0.5]{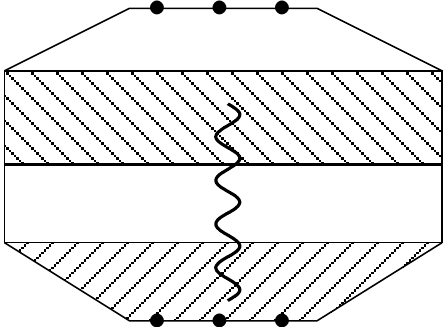}} &+&
\raisebox{-0.7cm}{\includegraphics[scale=0.5]{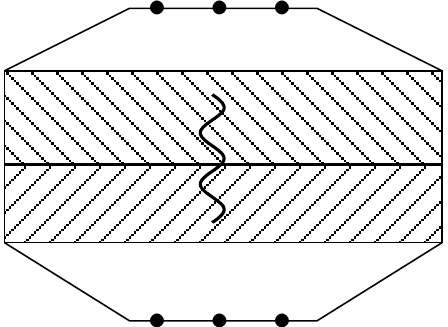}}\vspace{0.2cm} \\ 
C &=& C^{(1)} &+& C^{(2)} &+& C^{(3)} &+& C^{(4)} .
\end{array} \nonumber
\ee
On the left are the graphs contributing to the logarithm of the amplitude: The gluons couple to all edges of the two shaded regions
and the Lagrangian is inserted somewhere along the wigggly line.
(The figure includes a graph in which the two gluons are attached to the fermion propagator.)
On the right, each polygon is split into two, one part outside the box $(1,2,i{-}1,i)$ and one part inside, leading to four terms.
The reason we can do this so simply is that the gauge sector is essentially abelian.

Making sense of these figures requires specifying a path connecting $x_2$ to $x_{i{-}1}$.  Since these points are not null-separated, no path in spacetime will do
this in a conformal invariant fashion.  We want to preserve conformal invariance.  The way out is to use a path that goes through a fifth-dimensional AdS space.
Let us explain what this means.  It is here that the six-dimensional notation introduced above pays off.  The gauge link from two null-separated points $X_1$ and $X_2$ was written
in section \ref{sec:prelim} as
\be
 \frac{\l \theta X_1 X_2 I\theta'\r}{\l ZX_1\r \l ZX_2\r \l ZI\r}.
\ee
When we verified independence with respect to $I$ there,
nowhere did we need that $X_1$ and $X_2$ where null separated (nor even that they were points inside spacetime, $X_i^2=0$).  Thus
for general $X_1$ and $X_2$, we simply \emph{define} the gauge-link along an AdS geodesic to be that expression.  This provides a gauge-invariant
closure of the polygon.

\item We render the top and bottom polygons finite in the simplest possible way, by moving their endpoints slightly away from the corners of the box:
\be
\begin{array}{ccc}
\raisebox{-0.8cm}{\includegraphics[scale=0.8]{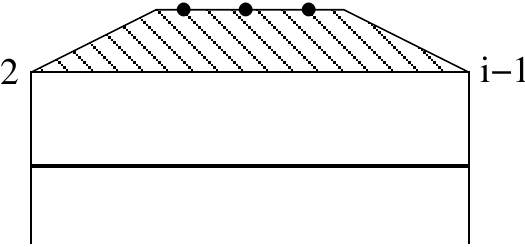} }&\to&
\raisebox{-0.8cm}{\includegraphics[scale=0.8]{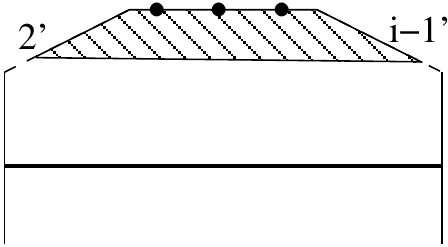} }.
\end{array} \nonumber
\ee
In equations,
\be
 Z_2\wedge Z_3 \to Z_{2'}\wedge Z_3 \equiv (Z_{2}-\epsilon_1 Z_{4})\wedge Z_{3}, \qquad
 Z_{i{-}1}\wedge Z_i \to Z_{i{-}1}\wedge Z_{i'} \equiv Z_{i{-}1} \wedge (Z_{i}-\epsilon_2 Z_{i{-}2}).
\ee 
The lower polygon is to be regulated in the same way.
$C^{(1)}$ is trivially finite as the regulators are taken away, but $C^{(2)}$ and $C^{(3)}$ diverge logarithmically with $\epsilon_1$ and $\epsilon_2$.
We will see, however, that these divergence can be subtracted and renormalized in a canonical and conformal invariant fashion.
Thus all infrared divergences will be shoved into $C^{(4)}$.

\item We would still have to regulate the graphs contributing to $C^{(4)}$, should we want to calculate them.  And we would have to carefully
carry out the subtraction of the two-loop correction to the cusp anomalous dimension that appear in the BDS Ansatz \cite{BDS}, following the definition of the remainder function.
However, by construction, $C^{(4)}_{i,j}$ will be a conformal invariant function of a \emph{single} cross-ratio $u_{i,j{-}1,j,i{-}1}$.
So, instead of computing it directly, we can use supersymmetry to obtain it from what we already have, through the sum rule (\ref{sumrule}).
This way, the whole computation is phrased directly in terms of the remainder function.
\end{itemize}

\subsection{Contribution $C^{(1)}$}

To obtain the integrand we begin by writing down the field strength induced by the top and bottom polygons
\ba
 (\theta G\theta')_\textrm{top}&=& \sum_{j=3}^{i-1} \frac{\l j{-}1jj{+}1\theta\r\l jI\theta'\r}{\l Zj{-}1j\r\l Zjj{+}1\r\l ZI\r} -
 \frac{\l \theta 23\, i{-}1i I\theta'\r}{\l Z23\r\l Zi{-}1i\r\l ZI\r}, \\
 (\theta G\theta')_\textrm{bottom}&=& \sum_{k=i{+}1}^{n{+}1} \frac{\l k{-}1kk{+}1\theta\r\l kI\theta'\r}{\l Zk{-}1k\r\l Zkk{+}1\r\l ZI\r}+
 \frac{\l \theta 12\, ii{+}1 I\theta'\r}{\l Z12\r\l Zii{+}1\r\l ZI\r}.
\ea
The spacetime point $Z$ will be the loop integration point and $\theta,\theta'$ are two arbitrary 4-spinors associated with that point.
We see two sorts of terms.  The terms inside the summation come from the null edges and are as in \Eq{inducedG}.
The extra term comes from the ``AdS edge'' mentioned above, and ensures that the field strength of each polygon is gauge invariant (independent of $I$).
We should remark that there is some freedom regarding the ordering of the first two factors in these terms, e.g. $XdX$ versus $-dX X$, which did not arise in the case of a null edge.
Any choice would be just as much gauge-invariant. The choice we have taken turns out to ensure the finiteness of the integrals, as will be fully apparent shortly.
It would be nice to understand better what is the special physical meaning of this choice.

The simplest way to compute the product $G_\textrm{top}G_\textrm{bottom}$ is to make the gauge choice $I=(ii{+}1)$
in both polygons.  This kills two edges in the bottom polygon and yields the integrand
\be
 \sum_{j=3}^{i{-}1}\sum_{k=i{+}2}^{n{+}1} I_5(i; j{-}1,j;k{-}1,k) 
 - \sum_{k=i{+}2}^{n{+}1} \int_Z \frac{\l Z (23i)\cap(\kbar)\r \l i{-}1ii{+}1k\r}{\l Z23\r\l Zi{-}1i\r\l Zii{+}1\r\l Zk{-}1k\r\l Zkk{+}1\r}.
\ee
This multiply the fermion exchange diagram computed in \Eq{1loopMHV}.
The integrand is easily verified to be finite, and can thus be evaluated without regularization.
Term by term it contains collinear singularities on edges $i$ and $i{+}1$, however.
What appears to be the simplest way to proceed is to regulate the individual terms by simply moving $(ii{+}1)\to (ii{+}1)+ \epsilon I'$.
The basic integrals that we need are
\ba
 I_5(i,j; I) &=& \int_{Z} \frac{ \l Z (i{-}1ii{+}1)\cap(j{-}1jj{+}1)\r\l I ij\r}{\l Zi{-}1i\r\l Zii{+}1\r\l Zk{-}1k\r\l Zkk{+}1\r\l ZI\r}
 \nl
&=& \phantom{+} \Li_2(1-u_{I,1,2,i}) - \Li_2(1-u_{,1,2,i{-}1}) + \Li_2(1-u_{2,i{-}1,i,1}) \nl &&
  + \Li_2(1-u_{X,i,i{-}1,1}) -\Li_2(1-u_{X,i,i{-}1,2})
 + \log u_{X,1,2,i}\log u_{X,i,i{-}1,1}
\label{defI5}
\ea
and
\ba
 I_4^{\textrm{2me}'}(i{-}1,i';j{-}1',j)&\equiv& \int_Z \frac{\l i{-}1i j{-}1j'\r\l i'i{+}1jj{+}1\r-\l i{-}1i jj{+}1\r\l i'i{+}1j{-}1j'\r}
 {\l Zi{-}1i\r\l Zi'i{+}1\r\l Zj{-}1j'\r\l Zjj{+}1\r}
\nl &=& 2\Li_2(1-v) + \log v \log w+ \mathcal{O}(\epsilon),
\ea
where $v=v_{i,j{-}1,j,i{-}1}$ and $w=u_{i',i{-}1,j,j{-}1'}\propto \epsilon^2$.
Indeed, using the Schouten identity
\be
 \l Z (23i)\cap(\kbar)\r \l \ibar k\r =   \l Z (\ibar)\cap(\kbar)\r \l 23ik\r + ( \l Z k{-}1k\r \l kk{+}1 (\ibar)\cap(i23)\r - (k{-}1\leftrightarrow k{+}1)),
\ee
everything is reduced to these two basic integrals.   The final result is written compactly
\be C_{2,i}^{(1)} = \log u_{2,i{-}1,i,1} \times \sum_{j=2}^{i{-}1} \sum_{k=i}^{n{+}1}
 \left[\Li_2(1-u_{j,k,(k{-}1),(j{+}1)}) +
  \log \frac{x_{j,k}^2}{x_{(j{+}1),k}^2} \log \frac{x_{j,k}^2}{x_{j,(k{-}1)}^2}\right], \label{c1}
\ee
if we use the notation that $x_{(j{+}1)}\equiv x_2$ when $j=i{-}1$, and $x_{(k{-}1)}\equiv x_1$ when $k=i$.  This is quite similar to the pairing of two Wilson loops in \cite{OPE3},
except now we have two edges in AdS.  (There is a slight abuse of notation here -- some of the dilogarithms and logarithms in the sum are divergent.  They can be removed
using the identity $\Li_2(1-u)=-\Li_2(1-1/u)-\frac12\log^2u$, at the cost of making the expression look less symmetrical.)

\subsection{Contributions $C^{(2)}$ and $C^{(3)}$}
\label{sec:c2}

To find this integrand, we now need the three point function for a field strength and two fermions.
The Feynman rules from \Eq{fermionaction} gives this as
\be
  V_{X,Y,Z} 
  = \int_{U} \frac{\l \theta U 2\r \l U\,(\ibar)\cap(I\theta')\r}{\l UX\r^2\l
  UY\r^2 \l UZ\r^2 \l ZI\r}.  \label{vertex0}
\ee
The integration over the spacetime point $U$ can be done using the standard Feynman parameterization procedure.
The six-dimensional notation offfers a particularly efficient way to achieve this; let us go through the various step in detail (see also \cite{masonskinner6d,drummondhennint,aldayint}).
One first combines denominators using Feynman parameters
\be
 \frac{1}{\l UX_1\r\l UX_2\r\cdots \l UX_n\r}= \Gamma(n)\int \frac{[d^{n{-}1}a_1\cdots a_n]}{\l U(a_1X_1+ a_2X_2 +\ldots + a_nX_n)\r^n}.
\ee
The measure $[d^{n{-}1}a_1\cdots a_n]$ is the projective measure, defined explicitly by setting one of the coordinates to 1 and integrating
over the $n{-}1$ others.  It does not matter which variable is set to 1.
In the present case, the loop integral that we need is the second derivative of the basic integral in \Eq{fundamentalint}:
\be
 \Gamma(6)\int_{CD} \frac{\l CD Y\r\l CD Z\r}{\l CDX\r^6} = \Gamma(4)\frac{\l XY\r\l XZ\r}{(\frac12 \l X X\r)^4}
 - \Gamma(3) \frac{\l YZ\r}{(\frac12\l XX\r)^3}.
\ee
Thus \Eq{vertex0} becomes
\ba
 V_{X,Y,Z}&=&\int \frac{[d^2abc]abc}{\l ZI\r} \left(\frac{
    6\l \theta (bY)2 \r\l (aX+cZ)\, (\ibar)\cap(I\theta')\r} {(ab \l XY\r + ac\l XZ\r + bc \l YZ\r)^4} -
  \frac{2\l \theta2 \, (\ibar)\cap(I\theta')\r}{(ab \l XY\r + ac\l XZ\r
    + bc \l YZ\r)^3} \right) \nl
 &=& \frac{1}{\l XY\r\l XZ\r\l YZ\r\l
  ZI\r} \left( \frac{\l \theta Y2\r\l X\, (\ibar)\cap(I\theta')\r}{\l
    XY\r} + \frac{\l \theta Y2\r\l Z\, (\ibar)\cap(I\theta')\r}{\l
    YZ\r} - \l \theta2 \, (\ibar)\cap(I\theta')\r\right)
\nl &=& 
\frac{\l \theta Y2\r\l \ibar\theta'\r}{\l XY\r\l XZ\r\l YZ\r^2}
+\frac{\l \ibar2\r\l \theta XYI\theta'\r}{\l XY\r^2\l XZ\r\l YZ\r\l ZI\r}.
\label{diagram1a}
\ea
In the last line we have used Schouten identities to rearrange the result to a nicer form.

As anticipated, we find a double pole in $\l YZ\r$ but only single poles in $\l XZ\r$.
We are going to perform the $\tau_Y$ integral first.  Before doing that, we should combine this term
with the segment from $Y$ to $X_{i{+}1}$, since this side of the square also has a nontrivial dependence on $Y$.
Parametrizing $Y=(Z_{i{-}1} - \tau_Y Z_{i{+}1})\wedge Z_i$, this gives
\ba
&& \int_0^\infty d\tau_Y\left( V_{X,Y,Z} + \frac{\l \ibar 2\r\l i{-}1i i{+}1\theta\r\l iI\theta'\r}{\l ZY\r\l Zii{+}1\r\l ZI\r\l XY\r^2}\right)
\nl 
&=&
\frac{\l \ibar \theta\r\l i{-}1i\theta'2\r}{\l Xi{-}1i\r\l XZ\r\l Zi{-}1i\r\l Zii{+}1\r}
-\frac{\l \ibar2\r\l \theta ii{+}1XI\theta'\r}{\l Xi{-}1i\r\l Xii{+}1\r\l XZ\r\l Zii{+}1\r\l ZI\r}.
\ea
The fact that this is rational lies at the heart of our approach.
The remaining two sides of the square are then obtained as in the previous section.
After adding them, the $I$ dependence disappears in a simple way, yielding finally
\be
 (\theta G\theta')_\textrm{bottom\, square} = \frac{\l \ibar \theta\r\l i{-}1i\theta'2\r}{\l Xi{-}1i\r\l XZ\r\l Zi{-}1i\r\l Zii{+}1\r}
  - \frac{\l \ibar2\r\l \theta 12ii{+}1X\theta'\r}{\l Xii{-}1\r\l Xii{+}1\r\l XZ\r\l Z12\r\l Z ii{+}1\r}.
\ee
We are nearly done with the computation of the integrand.  The final step is to contract this result
with $G_\textrm{top}$ which was obtained in the previous subsection.  Following the adopted strategy, this is to be done
with $(23)\to (23)'$ and $(i{-}1i)\to (i{-}1i)'$ in order to regulate collinear divergences.  Because everything is then safely finite, we can simplify our life by
making the gauge choice $I=(i{-}1i)'$ inside $G_\textrm{top}$, which removes a couple of terms.
A short computation using Schouten identities then reduces everything to the known integrals listed previously:
\be
 \int_1^2 d\tau_X \sum_{j=4}^{i{-}2} \left(\frac{\l 2i{-}1ij\r\l i\bar2\r}{\l Xi{-}1i\r\l Xij\r} I_5(X; i{-}1,i; j{-}1,j)
 + \frac{\l 2\ibar\r\l i\bar2\r}{\l Xi{-}1i\r\l Xii{+}1\r} I_5(i; 1,X; j{-}1,j)\right) + \mathcal{J},  \label{c20}
\ee
where
\begin{align}
\mathcal{J} = \int d_X\!\log\frac{\l Xi{-}1i\r}{\l X34\r} &\left(I_5(X; 2',3; i{-}1,i) -I^{\textrm{2me}'}(X,2';i{-}1',i)\right)
\nl
+ \int d_X\!\log\frac{\l Xi{-}1i\r}{\l Xii{+}1\r} &\left(I_5(i; 1,X; 2',3)+I_5(i; 1,X; i{-}2,i{-}1') +I^{\textrm{2me}'}(X,2';i{-}1',i)\right.
\nl &\phantom{(} \left.- I^{\textrm{2me}'}(1,2',i{-}1',i)\right).
\end{align}
This completes the computation of the integrand associated with the two-loop derivative $C^{(2)}$.
The next step is to renormalize this expression.  When the corners $2'$ and $i'$ of the polygon approach their physical values $2$ and $i$,
$\mathcal{J}$ diverges like
\begin{align}
 & \phantom{+} \log \l i{-}1i' ii{+}1\r \int_1^2  d_X\!\log\frac{\l Xi{-}1i\r}{\l X34\r} 
  \log u_{i,2,X,i{-}1}
 \nl
&+\log \l 122'3\r \int_1^2 d_X\!\log\frac{\l Xi{-}1i\r}{\l Xii{+}1\r} 
\log u_{i{-}1,1,X,3}.
\end{align}
These are all the divergences.
A useful fact, easily seen by integration by parts, is that these two integrals are equal, \emph{e.g.}, the logarithms combine into
$\log (\l i{-}1i' ii{+}1\r\l 122'3\r)$. This was guaranteed by symmetry.
Importantly, this means that the divergence can be renormalized in a simple and conformal invariant fashion,
by replacing this logarithm with $\log (\l 12i{-}1i\r\l23ii{+}1\r)$, which has the same little group weights.
We have thus succeeded in making this subset of the computation conformally invariant on its own.

We can do better.  As discussed in the preamble, in the case $i=5$ where the top polygon is a triangle, conformal symmetry ensures that 
$C^{(2)}_{2,5}$ depends only on $u_{1,4,5,2}$.
In general, it makes sense to improve our renormalization prescription by adding a counter-term depending only on $u_{2,i{-}1,i,1}$,
which means moving things between $C^{(2)}$ and $C^{(4)}$.  This allows us to choose a counter-term such
that the triangle vanishes.  Although we will not need it, we record for completeness the expression for the counter-term
\ba
 \mathcal{J}_\textrm{c.t.}&=&  -\int d_X\!\log\frac{\l Xi{-}1i\r}{\l X34\r} \log u_{i,2,X,i{-}1} \log u_{1,2',i{-}1',i}  \nl
 &&+ \int d_X\!\log\frac{\l Xi{-}1i\r}{\l Xii{+}1\r} \left( 2\Li_2(1-u_{i,1,2,i{-}1})-2\Li_2(1-u_{i,X,2,i{-}1})-\Li_2(1-u_{i,1,X,i{-}1}) \right.\nl
 && \hspace{2cm} \left. +\Li_2(1-u_{i,1,X,3}) - \Li_2(1-u_{i{-}1,1,X,3})+\log u_{i{-}1,1,X,3}\log u_{i,2,X,i{-}1}\right)
 \nl && +\int d_X\!\log\frac{\l Xi{-}1i\r}{\l X34\r}\left( \Li_2(1-u_{X,i,i{-}1,2}) + \log u_{X,i,i{-}1,2}\log u_{X,3,i{-}1,1}\right).
\ea
Apart from the first line this depends only on the cross ratio $u_{2,i{-}1,i,1}$.\footnote{
 This does not depend on the point $3$ because it enters only through
 the combination $\l X34\r/\l1234\r$.  In our parametrization of $X$, this is equal to $1$.}
What is more important, and can be obtained directly without working out $\mathcal{J}_\textrm{c.t.}$, is the renormalized expression:
\ba
 \mathcal{J}+\mathcal{J}_\textrm{c.t.} 
 &=& \phantom{+} \int d_X\!\log\frac{\l Xi{-}1i\r}{\l Xii{+}1\r} \left( \Li_2(1-u_{i,1,X,3}) - \Li_2(1-u_{i{-}1,1,X,3}) - \Li_2(1-u_{i,1,X,i{-}2})\right.
 \nl && \hspace{4cm} \left. +\Li_2(1-u_{i{-}1,1,X,i{-}2}) - \log u_{1,3,i{-}2,i}\log u_{i,1,X,i{-}1}\right)
 \nl && + \int d_X\!\log\frac{\l Xi{-}1i\r}{\l X34\r} \left( \Li_2(1-u_{2,i,i{-}1,3})-\Li_2(1-u_{X,i,i{-}1,3}) \right).
\ea
This vanishes for $i=5$.
Substituting this for $\mathcal{J}$ in \Eq{c20} and rearranging a telescopic sum, we conclude that
 \ba
 C^{(2)}_{2,i}&=&
 \phantom{+}\sum_{j=4}^{i{-}2} \int _1^2 d_X\!\log\frac{\l Xi{-}1i\r}{\l Xij\r} I_5(X;i{-}1,i;j{-}1,j)\nl
 && + \sum_{j=4}^{i{-}2} \int_1^2 d_X\!\log\frac{\l Xi{-}1i\r}{\l Xii{+}1\r} \left( \Li_2(1-u_{i{-}1,1,X,j}) + \Li_2(1-u_{i,j,j{-}1,1}) + \Li_2(1-u_{1,j,j{-}1,X})\right.
 \nl && \hspace{2cm}\left. -\Li_2(1-u_{i{-}1,1,X,j{-}1}) - \Li_2(1-u_{i,j,j{-}1,X}) + \log u_{i{-}1,1,X,j} \log u_{i,j,j{-}1,1}\right)
  \nl &&
 + \int_1^2 d_X\!\log\frac{\l Xi{-}1i\r}{\l X34\r} \left( \Li_2(1-u_{2,i,i{-}1,3})-\Li_2(1-u_{X,i,i{-}1,3})\right). \label{c2}
\ea

\subsection{Contribution $C^{(4)}$}

As explained above, we don't actually need to compute anything new to obtain $C^{(4)}$.
By construction, $C^{(4)}_{i,j}$ depends on a single cross-ratio and can be determined
through the sum rule (\ref{sumrule}).   Let us denote this cross-ratio as $u\equiv u_{i,j{-}1,j,i{-}1}$.
From considering the sum rule for the pentagon, where the remainder function and the contributions $C^{(1)},\ldots, C^{(3)}$ all vanish, we conclude
that $C^{(4)}(0)=0$.  In the hexagon case, the nonzero terms in the sum rule give
\be
 C_{2,5}^{(4)}(u) = -C_{2,5}^{(1)} - C_{2,6}^{(2)} - C_{2,4}^{(3)}. 
\ee
It is a nontrivial consistency check that the right-hand side depends only on the required cross-ratio; if that were not the case,
the proposal would be wrong.

In the hexagon case, \Eq{c1} evaluates to
\be
 C^{(1)}_{2,5} = \log u \times \left( \Li_2(1-u_{3,5,6,2})+\Li_2(1-u_{4,6,1,3}) + \log u_{3,5,6,2}\log u_{4,6,1,3} - \frac{\pi^2}{6}\right).
\ee
Representing $\log u$ as $\int_1^2 d_X\log \frac{\l X45\r}{\l X56\r}$, this can be added directly to the integral representation (\ref{c2}) of $C^{(2)}$ and $C^{(3)}$.
We find:
\ba \hspace{-1cm}
 C_{2,5}^{(4)}(u) &=& \int_1^2 \left[
 \begin{array}{l} \phantom{+}d_X\!\log \frac{\l X45\r}{\l X34\r} \left( \Li_2(1-u_{X,5,4,2})+\Li_2(1-u_{X,6,5,2})-\Li_2(1-u_{X,6,4,2})\right) \\
 + d_X\!\log \frac{\l X56\r}{\l X45\r} \left( \log u_{X,5,6,2}\log u_{X,3,4,1} \right) \\
 + d_X\!\log \frac{\l X61\r}{\l X56\r} \left( \Li_2(1-u_{X,4,5,1}) +\Li_2(1-u_{X,3,4,1})-\Li_2(1-u_{X,3,5,1}) \right)
\end{array} \right]\!\!.   
\ea
The three lines are actually all equal in magnitude, but alternating in sign --- this can be proved by integrating by part in the first or third line.
So the total is minus the second line. It can be expressed in terms of the required cross-ratio by a simple rescaling of $\tau_X$:
\ba
C_{2,5}^{(4)}(u) &=& \int_{\tau=0}^{\tau=\infty} d\!\left( \log \frac{\tau+u}{\tau+1}\right) \log(1+\tau) \log (1+ \frac{u}{\tau})
\nl &=& -2\Li_3(1-\frac{1}{u}) - \Li_2(1-\frac1u)\log u - \frac16 \log^3u + \frac{\pi^2}{6} \log u.  \label{c4}
\ea
The same expression (which is valid for $u>0$), with the appropriate cross-ratio, gives all $C_{i,j}^{(4)}$.

\subsection{Method for computing the symbol}
\label{sec:methodsymbol}

We still have to integrate \Eq{c2}.  The first step is to compute its symbol.  It is actually possible to do so without performing any integration.
We are not aware of any standard algorithm for doing so, but we will describe the method we have employed.
The method is based on computing discontinuities across branch cuts, and comparing with the discontinuities
of the iterated integral (\ref{iteratedintegral}).
The leading transcendentality branch cuts of the latter end at the zeros and poles of the leftmost entry of the symbol.
These discontuinuities themselves have branch points, which are at the zeros and poles of the second entry of the symbol.
Thus, by computing discontinuities of discontinuities and so on, one can read off the symbol.

We hope to elaborate elsewhere about the algorithm we have used to compute the discontinuities.
Let us just try outline the method for a one-dimensional integral such as \Eq{c2}.  Basically, there are exactly three phenomena to keep track of:
\begin{itemize}
\item A pole of the integrand makes a loop around an integration endpoint.
\item A branch cut endpoint of the integrand makes a loop around an integration endpoint.
\item The value of the integrand at an endpoint undergoes monodromy.
\end{itemize}
By keeping track of these phenomena, we can calculate the symbol of an arbitrary integral of the form \Eq{c2}.
The procedure is systematic and entirely algebraic, and in principle could be automated.
A simple yet nontrivial example illustrating it is given in Appendix \ref{app:procedure}.

We have applied this algorithm to the transcendentality-two integrand in \Eq{c2}.
At the worse point the procedure generated no more than order 50 terms for a given $j$. Still, several significant simplifications occured, and
the resulting symbol is reproduced in Appendix \ref{app:symbol}.  For the reader's convenience, it is also attached with this submission to the arXiv in the form of a Mathematica notebook.
Nicely, by adding suitable telescopic terms in $j$,
we found that the symbol can be expressed as a sum over $j$
of a general term $\Delta$, which is both integrable manifests the correct symmetry under parity combined with the reflection $2\leftrightarrow i$.

The parity-odd part of $\Delta$ turns out to match the 3-mass ``easy'' six-dimensional hexagon
computed recently in \cite{hexagon1,hexagon2,hexagon3,hexagon4} (this can also be seen directly from \Eq{c2}).
It would be nice to find a similar interpretation for the parity even part.

Let us summarize.  The differential of the remainder function is
\be
 d R_n = \sum_{i,j} C_{i,j} d\log \l i{-}1ii{+}1j\r  \label{maindiff2}
\ee
where
\ba
 C_{i,j} = C_{i,j}^{(1)} + C_{i,j}^{(2)} + C_{i,j}^{(3)} + C_{i,j}^{(4)}.
\ea
The functions $C^{(1)}$ and $C^{(4)}$ are given analytically in Eqs.(\ref{c1}) and (\ref{c4}).
The function $C^{(2)}$ admits the integral representation (\ref{c2}) and $C^{(3)}$ is obtained from it by symmetry.
We did not attempt to express $C^{(2)}$ explicitly in terms of polylogarithms, but its symbol is given in Appendix \ref{app:symbol}.

\section{Checks on the two-loop result}
\label{sec:checks}

In this section we discuss some consistency checks which, in our opinion, establish beyond reasonable doubt
that the obtained expression for the two-loop remainder function is correct.
 
\subsection{Hexagon case}
We have verified that the symbol given in Appendix \ref{app:symbol} reproduces that of the compact analytic formula given
in \cite{volovichetal} when $n=6$.

\subsection{Integrability}

The differential in \Eq{maindiff2} is only meaningful if it is consistent with $d^2=0$.  This is what we call the integrability condition.
We did not attempt to verify this at the level of the integral representation, but we did verify, using a computer, that it was true at the level of the symbol.
This establishes, at least, that the maximal functional transcendentality part of our differential is the differential of a function.
This property holds for the whole $C$, but not for the individual $C^{(m)}$.  It would be nice to understand how integrability
derives from the supersymmetry algebra in general.

\subsection{Physical discontinuities}

The discontinuities of a scattering amplitude have a clear physical interpretation.
For instance, the leading branch cuts should be unitarity cuts, and the location of a branch point should be some physical threshold.
In a theory with only massless particles, these thresholds occur when a separation
$(x_i-x_j)^2$ becomes null.  Therefore, the first entry of the symbol is expected to be always of the form $\l ii{+}1jj{+}1\r$.
We can keep going.  The branch cuts themselves have branch cuts.  While their physical interpretation is more delicate,
the kinematical considerations which dictate the location of the endpoints can involve at most one extra channel and therefore should
be similar from one- to two-loops.  The endpoints of higher discontinuities, however, may in principle depend on three or more channels at a time,
and could be very complicated conditions.

Thus, specifically the first two entries of the symbol are expected to contain only patterns of the form
which appear in a general 1-loop amplitude, like $\Li_2(1-u)$, or involve two independent cuts like $\log \l ii{+}1jj{+}1\r\log\l kk{+}1mm{+}1\r$.
(These expectations were also formulated in \cite{OPE3}.)
We believe that a general proof of this statement is possible by considering Feynman integrals \cite{nimaandI}.
Pleasingly, we find that our symbol fulfills precisely these expectations.

\subsection{Collinear limits}

The remainder function $R_n$ is designed to have simple collinear limits \cite{BDS}.  Let us consider the simplest such limit,
where two adjacent edges of the polygon become parallel.  In this limit the $n$-gon is reducing to a $(n{-}1)$-gon, losing three degrees of freedom.
The approach to the limit is therefore characterized by three parameters.
We find convenient to use the following parametrization of momentum twistors:
\be
 \hat Z_n = Z_{n{-}1}  + \epsilon (Z_n{-}2 + \tau Z_1)  + \epsilon' Z_2
\ee
where $\epsilon'\ll \epsilon$.  The parameters $\epsilon$ and $\epsilon'$ regulate the approach to the collinear limit, while $\tau$ describe the relative length of the collinear
segments $n$ and $n{-}1$.  For the remainder function, the dependence on all three parameters must drop out in the limit:
\be
 \lim_{\epsilon,\epsilon'\to 0} R_n(Z_1,\ldots,\hat Z_n) = R_{n{-}1}'(Z_1,\ldots,Z_{n{-}1}).
\ee
This imposes nontrivial constraints on the differential of $R$, namely
\be
 \lim_{\epsilon,\epsilon'\to 0} dR = dR'
\ee
which we now wish to verify. Our expression for the differential is a sum of terms of the form $C_{i,j}d\log \l i{-}1ii{+}1j\r$.
The log derivative of the small parameter $\epsilon'$ multiplies $(C_{n{-}1,1}+C_{n,n{-}2})$; this combination must vanish.
Similarly, from canceling the log derivative of $\epsilon$ and using the exact sum rule \Eq{sumrule}, we conclude that
$(C_{1,n{-}1}+C_{n{-}2,n})$ vanishes in the limit.  From the pole at $d\tau/\tau$ at $\tau=0$, which multiplies
$(C_{n{-}2,n}-C_{n{-}1,1})$, we conclude that both must vanish separately. Similarly from the pole at $\tau=\infty$. (Strictly speaking we can only conclude
that these quantities vanish separately when $\tau=0$ or $\infty$, but this provision does not seem to be realized for the 2-loop amplitude.)
There remains the finite parts to match on both sides, which produce a large number of relations which can be summarized very simply
\begin{subequations} \label{coll}
\ba
 \lim C_{i,j} &=& C'_{i,j}, \qquad i,j\neq n{-}1,n, \label{coll1} \\
  \lim C_{i,n{-}1}+C_{i,n} &=& C'_{i,n{-}1}, \qquad i \neq n{-}1,n, \label{coll2} \\
  \lim C_{n{-}1,j}+C_{n,j} &=& C'_{n{-}1,j}, \qquad j\neq n{-}1,n,
\ea
\end{subequations}
where the $C'$ pertain to the $(n{-}1)$-gon.  The extra conditions,
$\lim C_{n{-}1,1}=0$ and $\lim C_{n,n{-}2}=0$, can be viewed as special case of these.

As above, we have verified using a symbolic manipulation package that the limits (\ref{coll}) work out at the level of the symbol (up to $n=10$, which is generic).

\subsection{Two-dimensional kinematics}

As another important check on our results, we have compared with existing results in two-dimensional kinematics,
which were given for an arbitrary number of points in \cite{duhr2d,khoze2d}.
In this case we were able to fully check the integral representation, \Eq{c2}, not only its symbol.

In two-dimensional kinematics, the number of particles is taken to be even
and particles with odd and even labels are distinguished,
\be
 Z_{2i-1} \to \left(Z_{2i{-}1}^1, Z_{2i{-}1}^2,0,0\right),  \qquad
 Z_{2i} \to \left( 0,0, Z_{2i}^3, Z_{2i}^4 \right).  \label{2dcrossratios}
\ee
Four-brackets with two odd and two even labels factorize into products of two-brackets,
and all other four-brackets vanish. Provided that $C_{\textrm{even},\textrm{odd}}\to0$,
\Eq{maindiff} becomes
\be
 d R_n \to \sum_{\substack{i<j \\ \textrm{both odd}}} (C_{i,j} + C_{j,i})d\log \l ij\r +\sum_{\substack{i<j \\ \textrm{both even}}} (C_{i,j} + C_{j,i})d\log \l ij\r  \label{2ddiff}.
\ee

We would like to evaluate the functions $C_{i,j}$ in these kinematics.  First of all, the vanishing of  $C_{\textrm{even},\textrm{odd}}$ is easy to see.
For odd $i$, we have that:
\be
 u_{2,i{-}1,i,1} \to \frac{\l 2i{-}1\r\l 2i{+}1\r\l 1i\r\l 3i\r}{\l 2i{-}1\r\l 2i{+}1\r\l 1i\r\l 3i\r} =1
\ee
form which $C^{(1)}$ and $C^{(4)}$ trivially vanish. In two-dimensional kinematics, we write $X=Z_2\wedge A$ where $A=Z_1-\tau Z_3$
in \Eq{c2}.  It is then also easy to see that $C^{(2)}_{2,i}$ vanishes for $i$ odd, because the measure becomes $d_X\log \frac{\l Xi{-}1i\r}{\l Xii{+}1\r}\to d_A\log \frac{\l Ai\r}{\l Ai\r}=0$.

Thus $C_{2,i}$ is only nonzero for even $i$.  These are slightly messy to compute,
because the way we have organized our computation is not really tailored for the special simplifications that occur in the special kinematics.
We give some details of this computation in Appendix \ref{app:2d}.
In the case of the octagon, the final result of that Appendix is simply
\ba
 C_{2,4} &=& -\log u_{1,3,5,7} \log u_{3,5,7,1} \log u_{4,6,8,2}, \nl
 C_{2,6} &=& \log u_{1,3,5,7} \log u_{3,5,7,1} \log (u_{2,4,6,8} u_{4,6,8,2}) \nonumber
\ea
plus cyclic.  Equation (\ref{2ddiff}) can then be easily integrated:
\be
 R_8 = -2\log u_{1,3,5,7} \log u_{3,5,7,1} \log u_{2,4,6,8} \log u_{4,6,8,2} + \textrm{Constant}.
\ee
This agrees precisely with \cite{duhr2d}.\footnote{We are factoring out a 2-loop prefactor $g^4$ where $g^2=\frac{g^2_\textrm{YM}N_c}{16\pi^2}$,
so our $R_n$ is four times that of \cite{duhr2d}.}
The constant cannot be determined by our methods; its value, $\textrm{Constant}=-\frac{2\pi^4}{9}$, may be fixed from triple collinear limits and the hexagon \cite{duhr2d}.\footnote{The author thanks Claude Duhr and Vittorio Del Duca for explanations regarding this point.}

\section{Conclusions}
\label{sec:conclu}

We have considered, in planar $\mathcal{N}=4$ super yang-Mills, a generalization of scattering amplitudes
and conjectured that it obeys simple Ward identities under dual superconformal symmetries.
These Ward identities follow from naive field theory manipulations, and the non-obvious part of the conjecture is that they should not be spoiled by any sort of divergences,
which arose previous treatments of superconformal symmetry on scattering amplitudes.
Somewhat analogously to the Ward identities employed in the classic work of Parke and Taylor \cite{parketaylor} on MHV amplitudes nearly three decades ago,
this allows a difficult computation involving gluons to be replaced with a less difficult ones involving fermions.
In the present case, the computation with fermions is related to a derivative of the computation with bosons.

Exploiting these Ward identities we obtained an integral representation for
the total differential of the two-loop general $n$-point MHV remainder function.  The result agrees with all
previously known analytic two-loop results, and extends them. Its symbol is reproduced in Appendix \ref{app:symbol}.
We have really only used $\mathcal{N}=1$ supersymmetry, and our results are thus valid for null polygonal Wilson loops in any $\mathcal{N}=1$ superconformal theory.
The next natural step would be to integrate this expression.  It would also be worthwhile to compare it numerically with the results \cite{numerical}.
With some work, we believe the method could be realistically expected to generate new results for three-loop MHV or two-loop NMHV amplitudes \cite{6ptNMHV}.

The generalization is based on the introduction of new variables which render exact all the dual supersymmetries,
e.g. those which act naturally on Wilson loops.  This includes a chiral half of the spacetime superconformal symmetries.
An important question, which remains to be addressed, is whether the other chiral half also becomes exact.
This would imply exact Yangian symmetry for the same price.
This would also suggest a generalization for scattering amplitudes in superconformal theories other than $\mathcal{N}=4$.

The methods we have used should commute with the Wilson loop OPE \cite{OPE1}, and with structural constrains derived from it \cite{severvieiratoappear}.
It might be fruitful to combine the two approaches.

One may ask whether general constraints can be derived regarding the structure of amplitudes, coming from the existence of this new object.
Our results provide evidence that this is the case.
Of course, in non-chiral superspace any function can be supersymmetrized, so supersymmetry alone does not provide much of any constraint.
However, we have found empirically that individual $\bar\theta,\theta$ components of the amplitude at two-loop have unexpectedly simple leading singularities,
all equal to ``1'' in the natural normalization, which in turns implies a very simple form for the last entry of the symbol at the same loop order.
This property begs for an explanation, which we hope will help unlock the structure of the symbol at higher loops.


\acknowledgments{
It is a pleasure to acknowledge useful discussions with
Christian Vergu, Niklas Beisert, Dave Skinner, Yutin Huang, Nima Arkani-Hamed, Jake Bourjaily, Juan Maldacena,
Tristan McLoughlin, Amit Sever, Pedro Vieira and Gregory Korchemsky.
Part of this work was completed at the Kavli Institute for Theoretical Physics during the program ``The harmony of scattering amplitudes'',
which the author thanks warmly for hospitality.  The author gratefully acknowledges support from NSF under grants PHY-0969448 and PHY05-51164.
}

\appendix
\pagebreak
\section{Symbol of $C_{i,j}$}
\label{app:symbol}

The differential of the n-point function is expressed as
\be
 d R_n = \sum_{i,j} C_{i,j} d\log \l i{-}1ii{+}1j\r
\ee
where $C_{2,i}$ is the sum of the four contributions
\ba
\label{symbol}
 C_{2,i}^{(1)}&=& \log u_{2,i{-}1,i,1} \times \sum_{j=2}^{i{-}1} \sum_{k=i}^{n{+}1}
 \left[\Li_2(1-u_{j,k,k{-}1,j{+}1}) +
  \log \frac{x_{j,k}^2}{x_{j{+}1,k}^2} \log \frac{x_{j,k}^2}{x_{j,k{-}1}^2}\right], \nl
 C_{2,i}^{(2)} &=& \sum_{j=4}^{i-2} \Delta(1,2; j{-}1,j; i{-}1,i), \nl
 C_{2,i}^{(3)} &=& \sum_{j=i{+}2}^{n} \Delta(2,1; j,j{-}1; i,i{-}1), \nl
 C_{2,i}^{(4)} &=&  -2\Li_3(1-\frac{1}{u}) - \Li_2(1-\frac1u)\log u - \frac16 \log^3u + \frac{\pi^2}{6} \log u,
\ea
and other $C_{i,j}$ are obtained by cyclic symmetry.
In the first line, $x_{j{+}1}\equiv x_2$ when $j=i{-}1$, and $x_{k{-}1}\equiv x_1$ when $k=i$, and in the last line, $u=u_{2,i{-}1,i,1}$.
The symbol of $\Delta$ is
\begin{align}
& \SS \Delta(1,2; j{-}1,j; i{-}1,i) \nl
 &=  \left(\phantom{+}\SS[I_5(i; 1,2; j{-}1,j)] \otimes \frac{ \l ii{+}1(\bar2)\cap(\jbar)\r\l 23ij\r}{\l j{-}1jj{+}1i\r\l 123j\r\l23ii{+}1\r}  - ((ii{+}1)\to (i{-}1i))\right) \nl  
 & \hspace{-1cm} +\left( \begin{array}{l}\phantom{+}
 \frac12\SS[\Li_2(1-u_{j,2,1,i{-}1})-\Li_2(1-u_{j,2,1,i}) ] 
 \otimes
 \left(\frac{\l123i\r\l j{-}1jj{+}12\r\l 23ij\r}{\l 123j\r\l j{-}1jj{+}1i\r\l23ii{+}1\r}\right)^2
 \frac{\l jj{+}1(\bar2)\cap(\ibar)\r\l ii{+}1jj{+}1\r}{\l 2ijj{+}1\r\l 13(2i{-}1i)\cap(2jj{+}1)\r}\\
   + \frac12\SS[\Li_2(1-u_{j,i{-}1,i,2})-\Li_2(1-u_{j,i{-}1,i,1})]\otimes 
   \left(\frac{\l 12i{-}1i\r\l 23ij\r}{\l 123i\r\l i{-}1ii{+}1j\r\l 23i{-}1i\r}\right)^2
   \frac{\l jj{+}1(\bar2)\cap(\ibar)\r \l i{-}1i{+}1(i23)\cap(ijj{+}1)\r}{\l 2ijj{+}1\r\l 12jj{+}1\r}\\
   +\frac12\SS[\Li_2(1-u_{2,i{-}1,i,1})] \otimes \frac{\l jj{+}1(\bar2)\cap(\ibar)\r\l i{-}1i{+}1(i23)\cap(ijj{+}1)\r}{\l 2ijj{+}1\r\l 13(2i{-}1i)\cap(2jj{+}1)\r}\\
   +  \frac12\SS[\log u_{j,i{-}1,i,2}\log u_{j,2,1,i{-}1}]\otimes \left(\frac{\l23ij\r}{\l 123j\r}\right)^2 \frac{\l jj{+}1(\bar2)\cap(\ibar)\r\l 13(2i{-}1i)\cap(2jj{+}1)\r}{\l 2ijj{+}1\r\l23i{-}1i\r\l i{-}1i{+}1(i23)\cap(ijj{+}1)\r} \\
   \quad - ((jj{+}1)\to (j{-}1j))
   \end{array}
  \right)\nl
 &+ \SS[I_5(1; i{-}1,i; j{-}1,j)] \otimes \frac{\l 12ij\r\l23i{-}1i\r}{\l 12i{-}1i\r\l 23ij\r} \nl
  &+ \SS[\log u_{i,j{-}1,j,1}\log u_{2,i{-}1,i,1}] \otimes \frac{\l j{-}1j{+}1 (j12)\cap(jii{+}1)\r\l123i\r\l23i{-}1i\r}{\l123j\r\l j{-}1jj{+}1i\r\l 12i{-}1i\r\l23ii{+}1\r}.
\end{align}
The factors of $\frac12$ cancel telescopically in the sum over $j$, and there are no $\frac12$ in front of anything in the full symbol of the amplitude (e.g., inside the big parenthesis,
only the squared factors do not telescope away).
The symbol could be written more succintly by exploiting these telescopic cancellations; this particular presentation
makes the individual term $\Delta$ integrable and parity covariant. $I_5$ is the ``pentagon integral''
\ba \hspace{-1cm}
 I_5(X; 1,2; i{-}1,i)&=& \phantom{+} \Li_2(1-u_{X,1,2,i}) - \Li_2(1-u_{X,1,2,i{-}1}) + \Li_2(1-u_{2,i{-}1,i,1}) \nl &&
  + \Li_2(1-u_{X,i,i{-}1,1}) -\Li_2(1-u_{X,i,i{-}1,2})
 + \log u_{X,1,2,i}\log u_{X,i,i{-}1,1}.
\ea

\section{Procedure for computing the symbol}
\label{app:procedure}

In this section we wish to illustrate the method  for computing the symbol outlined in subsection \ref{sec:methodsymbol}, on a simple example;
we hope to expand more on the method elsewhere.
\be
 I_\textrm{example} = \int_0^\infty \frac{dx (a-b)}{(x+a)(x+b)} \log(1+ x) = \Li_2(1-b) - \Li_2(1-a). \label{explicitex}
\ee
We want to compute its symbol without actually doing the integral.
The first step in the method is to find a function dual to the boundaries of the integration region: $f(x)=x/d$, where $d$ is an arbitrary nonzero constant.
This function is characterized by having a pole on the endpoint at infinity and a zero at the other endpoint.
The three phenomena referred to in the main text then contribute as follow.
The first phenomenon always contribute, for a pole located at point $-a$, an amount $f(-a)$ tensored with the residue of the pole:
\be
 \frac{a}{d} \otimes (1-a).  \label{eqB2}
\ee
(Signs can be dropped for entries inside the symbol.) The pole at $x=b$ contributes
\be
- \frac{b}{d} \otimes (1-b). \label{eqB3}
\ee
The second phenomenon contributes $f(-1)$ tensored with an integration which ends at the branch point $1$:
\be
 \frac{1}{d} \otimes \SS\int^1 \frac{dx (a-b)}{(x+a)(x+b)}.
\ee
The third phenomenon replicates an existing endpoint, and moves the leftmost entry of the symbol of the integrand, evaluated at the endpoint, outside the integral:
\be
 (1-0) \otimes \SS\int_0 \frac{dx (a-b)}{(x+a)(x+b)}.
\ee
Note that in this case the first entry is 1, so this contribution vanishes. There is also the endpoint at infinity.
In the present example, phenomena 2 and 3 occur simultaneously at that point.
A simple and systematic way to untangle this, is to separate the two endpoints
by chopping the integral at $\Lambda$, so that $f=x\Lambda/d(x-\Lambda)$, and then take $\Lambda\to\infty$:
\be
 (1-\Lambda) \otimes \SS \int^\Lambda  \frac{dx (a-b)}{(x+a)(x+b)}
- \frac{\Lambda}{d} \otimes \SS \int^\infty \frac{dx (a-b)}{(x+a)(x+b)}
  \Rightarrow d \otimes \SS \int^\infty \frac{dx (a-b)}{(x+a)(x+b)}.
\ee
The integrals we have written down have only one endpoint and, correspondingly, would seem to not make sense.
A nontrivial fact, which will always be guaranteed, is that the endpoints organize into pairs.
Here the contributions of the second and third type add up to
\be
 d\otimes  \SS \int_1^\infty \frac{dx (a-b)}{(x+a)(x+b)} = d \otimes \frac{1-a}{1-b}.
\ee
A second nontrivial fact, which can be used either as a check on the computation, or as a way to remove all but (\ref{eqB2}) and (\ref{eqB3}),
is that the dependence on $d$ cancels in the total:
\be
 \SS [I_\textrm{example}] = a \otimes (1-a) - b\otimes (1-b)
\ee
in agreement with the explicit result of doing the integral.

\section{Two-dimensional kinematics}
\label{app:2d}

In this section we complete the computation of the function $C_{i,j}$ in the kinematics (\ref{2dcrossratios}).
The function $I_5(X;i{-}1,i; j{-}1,j)$ that enters \Eq{c2} vanishes when both $i$ and $j$ are even (this can be seen directly, pre-integration,
from the vanishing of the numerator in \Eq{defI5}).  For $j$ odd it is pure logarithms.
(Passing from forms with dilogarithms to forms with logarithms is easy, if one uses the symbol, and fixes $\pi^2/6$ ambiguities by using special points.)
After collecting some telescopic terms in \Eq{c2} we find ``only''
\ba
 C_{2,i}^{(2)} &=& \phantom{+}\sum_{j=5\,\textrm{odd}}^{i{-}3} \int_1^3 d_A\!\log\frac{ \l Ai{-}1\r}{\l Aj\r} 
 \log \frac{\l Ai{-}1\r\l i{+}1j\r}{\l Ai{+}1\r\l i{-}1j\r} \log \frac{\l2j{+}1\r\l j{-}1i\r}{\l2j{-}1\r\l j{+}1i\r}
 \nl
 && + \sum_{j=5\,\textrm{odd}}^{i{-}3} \int_1^2 d_A\!\log\frac{\l Ai{-}1\r}{\l Ai{+}1\r} 
 \log \frac{\l Ai{-}1\r\l j1\r}{\l Aj\r\l i{-}11\r} \log \frac{\l2j{+}1\r\l j{-}1i\r}{\l2j{-}1\r\l j{+}1i\r}
 \nl
 && + \int_1^3 d_A\!\log \frac{\l Ai{-}1\r}{\l Ai{+}1\r}
  \left( 
 \Li_2(\frac{\l 1A\r\l i{+}13\r}{\l i{+}1A\r\l13\r})-\Li_2(\frac{\l 1A\r\l i{-}13\r}{\l i{-}1A\r\l13\r})-\Li_2(\frac{\l 1A\r\l i{+}1i{-}1\r}{\l i{+}1A\r\l1i{-}1\r})\right.
 \nl && \hspace{0cm} \left.
 +\log \frac{\l Ai{+}1\r\l i{-}11\r}{\l Ai{-}1\r\l i{+}11\r}\log \frac{\l i{+}1i{-}1\r\l 13\r}{\l i{+}13\r\l 1i{-}1\r}\right)
 - \int_1^3 d_A\!\log \frac{\l Ai{-}1\r}{\l A3\r} \Li_2(\frac{\l A3\r\l i{-}1i{+}1\r}{\l Ai{-}1\r\l 3i{+}1\r}). \nonumber
\ea
This looks a mess but it is not so bad. The first two lines sum up to simple logarithms.
The remaining dilogarithms are a bit vexing, but they can be removed by a simple integration by parts; this sort of simplification is specific to the special kinematics.
Therefore
\ba
 C_{2,i}^{(2)} &=& 
 \phantom{+} \sum_{j=5\,\textrm{odd}}^{i{-}3} \log \frac{\l 3i{-}1\r\l i{+}1j\r}{\l 3i{+}1\r\l i{-}1j\r}\log \frac{\l 3i{-}1\r\l j1\r}{\l 3j\r\l i{-}11\r}\log \frac{\l2j{+}1\r\l j{-}1i\r}{\l2j{-}1\r\l j{+}1i\r}
 \nl && 
 + \int_{\tau=0}^{\tau=\infty} d\log \frac{\tau+u}{\tau+1} \left( \log \frac{\tau+u}{\tau+1}\log \frac{1-u}{1+\tau}-\frac12 \log^2(1+\tau) \right). \label{dummyc2}
\ea
The second line is still a bit ugly.  However, beautiful things happen when we add to it $C^{(3)}$, which contains the same term,
and $C^{(4)}$: the sum is simply $\log u\Li_2(u)$.  This cancels a term from $C^{(1)}$, which we can then write compactly
\ba
\hspace{-1cm} C^{(1)}_{2,i} + \log u \Li_2(u) \!&=&\! \log u \!\times\! \left( \begin{array}{l}
\phantom{+} \sum_{j=5\,\textrm{odd}}^{i{-}1} \sum_{k=i{+}2\,\textrm{even}}^{n} \log u_{3,k{+}1,k{-}1,j} \log u_{i,j{-}1,j{+}1,k} \\
+\sum_{j=4\,\textrm{even}}^{i{-}2} \sum_{k=i{+}3\,\textrm{odd}}^{n{+}1} \log u_{2,k{+}1,k{-}1,j}\log u_{i{+}1,j{-}1,j{+}1,k} 
\\+\log(1-u)\log u_{2,2,i,i}
\end{array}\right) \!.\label{lastexpr}
\ea
The cross-ratio $u_{2,2,i,i}$ is not a typo, it means we are somewhat abusing notation to produce a more compact expression:
the ill-define brackets $\l22\r$ cancels out e.g. between the second line with $k=n{+}1$ and the third line.
Our final result for $C_{2,i}$ is the first line of \Eq{dummyc2}, plus an equivalent term for $C^{(3)}$, plus \Eq{lastexpr}.
For the octagon this is discussed in the main text.  We have verified symbolically that twice this result agrees
with the derivative with respect to $\l 2i\r$ of the general formula in \cite{khoze2d} (up to $n=20$, which is amply generic),
confirming the results of section \ref{sec:main} in the case of special kinematics.

\end{document}